\documentclass[12pt, a4paper]{article}
\pdfoutput=1
\usepackage{graphicx,amsmath,amssymb,booktabs,enumerate}
\usepackage{subfigure, url}
\usepackage{comment}
\usepackage{color}

\def\beq{\begin{eqnarray}}
\def\eeq{\end{eqnarray}}

\def\SU{\mathop{\rm SU}}

\allowdisplaybreaks[1]

\newcommand{\1}{\mbox{1}\hspace{-0.25em}\mbox{l}}

\usepackage[colorlinks=true, linkcolor=blue, citecolor=blue,
urlcolor=blue]{hyperref} 

\begin{document}
\baselineskip 0.7cm
\begin{titlepage}

\begin{flushright}
\end{flushright}

\vskip 1.35cm
\begin{center}
{\Large \bf 
Sfermion Flavor and Proton Decay \\in High-Scale Supersymmetry
}
\vskip 1.2cm

Natsumi Nagata${}^{1}$ and
Satoshi Shirai${}^{2}$
\vskip 0.4cm

{\it
$^1$ Department of Physics,
Nagoya University, Nagoya 464-8602, Japan\\
and Department of Physics, 
University of Tokyo, Tokyo 113-0033, Japan\\
$^2$Berkeley Center for Theoretical Physics, Department of Physics, \\
and Theoretical Physics Group, Lawrence Berkeley National Laboratory, \\
University of California, Berkeley, CA 94720, USA \\
}

\vskip 1.5cm

\abstract{
 The discovery of the Higgs boson with a mass of around 125 GeV gives a
 strong motivation for further study of a high-scale supersymmetry
 (SUSY) breaking model. In this framework, the minimal SUSY SU(5) grand
 unification model may be viable since heavy sfermions suppress the
 proton decay via color-triplet Higgs exchanges. At the same time,
 sizable flavor violation in sfermion masses is still allowed by
 low-energy precision experiments
 when the mass scale is as high as ${\cal O}(100)$~TeV, which naturally
 explains the 125~GeV Higgs mass. In the presence of the sfermion flavor
 violation, however, the rates and branching fractions of proton decay
 can be drastically changed. In this paper, we study the effects of
 sfermion flavor structure on proton decay and discuss the experimental
 constraints on sfermion flavor violation. We find that proton-decay
 experiments may give us a valuable knowledge on sfermion flavor
 violation, and by combining it with the results from other low-energy
 precision experiments, we can extract insights to the structure of
 sfermion sector as well as the underlying grand unification model. 
}
\end{center}
\end{titlepage}
\setcounter{page}{2}


\section{Introduction}

A high-scale supersymmetry (SUSY) breaking model
\cite{Wells:2003tf,ArkaniHamed:2004fb, Giudice:2004tc,
ArkaniHamed:2004yi,Wells:2004di}, in which the sfermion mass scale is
much higher than the weak scale, has many attractive features from
various points of view, such as the SUSY flavor/CP problems and the
cosmological problems. In particular, the discovery of the Higgs boson
with a mass of around 125~GeV \cite{Aad:2012tfa, Chatrchyan:2012ufa},
which is somewhat too heavy for a weak-scale minimal SUSY standard model
(MSSM)  \cite{OYY,Giudice:2011cg}, seems to give the strongest
motivation for the high-scale SUSY model. For this reason, both
theoretical and phenomenological aspects of such a framework have been
further investigated, especially after the Higgs discovery
\cite{spread,PGMs,Arvanitaki:2012ps,ArkaniHamed:2012gw}.

Such a scenario is also helpful for the construction of a grand
unification theory (GUT). Decoupling sfermions does not affect the
successful gauge coupling unification at one-loop level, since they form
complete SU(5) multiplets. Indeed, 
the unification can be improved in a sense, as the threshold
corrections to the gauge couplings at the GUT scale can be small
compared with the low-scale SUSY ones \cite{Hisano:2013cqa}. In
addition, heavy sfermions prevent too rapid proton decay
\cite{Murayama:2001ur} via the dimension-five operators $QQQL/M_{\rm
GUT}$ and $\bar{u}\bar{e}\bar{u}\bar{d}/M_{\rm GUT}$ 
generated from the color-triplet Higgs exchanges. Recently, the proton
decay in the minimal SUSY $\SU(5)$ GUT was reexamined and it was shown
that ${\cal O}(100)$~TeV sfermions, which explain the 125~GeV Higgs
mass, can be consistent with the current constraints
\cite{Hisano:2013exa}.  

However, it was also pointed out that Planck-suppressed operators
$QQQL/M_P$ and $\bar{u}\bar{e}\bar{u}\bar{d}/M_{P}$
with ${\cal O}(1)$ coefficients result in 
too rapid proton decay even in the high-scale SUSY model
\cite{Dine:2013nga}. This discrepancy clearly comes from the underlying
assumptions of a flavor symmetry. The operators from the color-triplet
Higgs exchanges are suppressed by small Yukawa couplings. The flavor
symmetry which realizes the Yukawa hierarchy may reduce the coefficients
of such Planck-suppressed operators.

Even if such a flavor symmetry actually exists and the dangerous
dimension-five operators are well suppressed, the sfermion flavor
structure is not necessary under control.
This is because the flavor charges of non-holomorphic
operators like $Q_i^{} Q_j^{\dagger}$, which relate to soft sfermion
masses, depend on the underlying models. Therefore, large flavor
violation in the sfermion masses may occur in some flavor
models. In fact, such sizable flavor violation
can be allowed in the high-scale SUSY scenario; if the sfermion mass
scale is much larger than 100~TeV, even the maximal flavor violation may
be consistent with the current experimental constraints
\cite{McKeen:2013dma,Sato:2013bta,Altmannshofer:2013lfa}.

The sfermion structure considerably affects the proton decay rate. In
the previous study \cite{Hisano:2013exa}, however, such effects are not
considered. Since sizable flavor violation may be present in the case of
high-scale SUSY, it is important to find out the consequence of
flavor violation on proton decay and to examine it in proton-decay
experiments. In this paper, therefore, we study the impact of the
sfermion flavor structure on the proton decay in the minimal SU(5) GUT
model with high-scale SUSY. It is found that the resultant proton decay
rate is drastically changed depending on the sfermion flavor structure,
which gives strong constraints on the flavor violation in the
sfermion sector. Further, we will find a smoking-gun signature for the
sfermion flavor violation, which may be searched in future proton-decay
experiments. In consequence, proton-decay experiments might shed light
on the structure of sfermion sector even when the SUSY scale is much
higher than the electroweak scale. 

This paper is organized as follows: in the next section, we introduce a
high-scale SUSY model which we deal with in the following discussion, and
give a brief review of the current experimental constraints on flavor
violation in the sfermion sector. Then, in Sec.~\ref{sec:protondecay},
we evaluate the proton decay rates in the presence of sfermion
flavor violation and discuss the experimental bounds on
it. Section~\ref{sec:summary} is devoted to summary and discussion.

\section{High-Scale SUSY Model}

\subsection{Mass Spectrum}

To begin with, let us briefly discuss a high-scale SUSY model which we
consider in the following discussion. Suppose that the supersymmetry
breaking field $X$ is charged under some symmetry. This suppresses the
operators linear in $X$ but allows $X^\dagger X$ couples to the MSSM
superfields. Especially, the following terms in the K\"{a}hler potential
can be present:
\begin{equation}
  K \ni -\frac{c}{M_*^2} X^\dagger X \Phi_{\rm MSSM}^\dagger \Phi_{\rm
   MSSM} ^{}~,
\label{eq:scalar}
\end{equation}
where $\Phi_{\rm MSSM} = \Phi_M, H_u, H_d$, and $c$ is an ${\cal O}(1)$
parameter, which depends on the species. $M_*$ is the cutoff scale of
the theory. These terms give soft masses as $m_0^2 = c|F_X|^2/M_*^2$ for
the MSSM scalars, with $F_X$ the $F$-term vacuum expectation value (VEV)
of the field $X$. One of the natural choices of $M_*$ is the Planck
scale $M_P$. In this case, $m_0$ is almost the same as the gravitino mass
$m_{3/2}$. 
 
The supersymmetric Higgs mass $\mu_H$ and the soft $b$-term may be
generated via
\begin{equation}
K \ni -\frac{c^\prime}{M_*^2} X^{\dagger} X H_u H_d + c^{\prime
 \prime} H_u H_d + {\rm h.c.}, 
\end{equation}
which leads to $b = c^\prime |F_X|^2/M_*^2 + c^{\prime
\prime}|m_{3/2}|^2$ and $\mu_H = c^{\prime \prime} m_{3/2}^*$
\cite{Giudice:1988yz, Inoue:1991rk, Casas:1992mk}. 
Because of the charge of the SUSY breaking filed $X$, direct couplings
of $X$ to the gauge supermultiplets and the superpotential can be
forbidden by the symmetry. The main contribution to the gaugino masses
and the trilinear $A$-terms in this case arises from the anomaly
mediation effects. With pure anomaly mediation effects \cite{AMSB}, the
gaugino masses are given by
\begin{eqnarray}
  M_{\tilde B} = \frac{3}{5} \frac{11 \alpha_1}{4\pi} m_{3/2},~~~
  M_{\tilde W} = \frac{\alpha_2}{4\pi} m_{3/2},~~~
  M_{\tilde g} = \frac{-3 \alpha_3}{4\pi}  m_{3/2}~,
\end{eqnarray}
where $\alpha_a\equiv g_a^2/4\pi$ $(a=1,2,3)$ and $M_a$ $(a=\tilde
B,\tilde W,\tilde g)$ are the gauge 
couplings and the gaugino masses, respectively.
This mass relation can be modified via quantum corrections from the SUSY
breaking effects by the MSSM particles \cite{Pierce:1996zz} or extra
particles in some higher-energy scale \cite{Nakayama:2013uta,
Harigaya:2013asa}.  
The trilinear $A$-terms are also suppressed by a loop-factor and thus we
neglect them hereafter.

Next, we introduce our convention for the sfermion mass-squared
matrices. The soft mass terms of sfermions are given as
\begin{equation}
 {\cal L}_{\rm soft}=
-\widetilde{Q}^*_{Li}({m}^2_{\tilde{Q}_L})_{ij}\widetilde{Q}_{Lj}
-\widetilde{L}^*_{Li}({m}^2_{\tilde{L}_L})_{ij}\widetilde{L}_{Lj}
-\widetilde{u}_{Ri}^{*}({m}^2_{\tilde{u}_R})_{ij}\widetilde{u}_{Rj}^{}
-\widetilde{d}_{Ri}^{*}({m}^2_{\tilde{d}_R})_{ij}\widetilde{d}_{Rj}^{}
-\widetilde{e}_{Ri}^{*}({m}^2_{\tilde{e}_R})_{ij}\widetilde{e}_{Rj}^{}
~,
\end{equation}
where $i,j=1,2,3$ denote the generation indices. The squark mass
matrices are defined in the so-called super-CKM basis, in which the
up-type quark mass matrices are diagonal and squarks are rotated in parallel to
their superpartners.
We further parametrize their structure as follows:
\begin{equation}
{m}^2_{\tilde{f}} \!=\! {m}_0^2 
\left( 
\begin{matrix}
1+\Delta_1^{\tilde{f}} & \delta_{12}^{\tilde{f}} & \delta_{13}^{\tilde{f}}\\
\delta_{12}^{\tilde{f}*} & 1+\Delta_2^{\tilde{f}} & \delta_{23}^{\tilde{f}}\\
\delta_{13}^{\tilde{f}*}  & \delta_{23}^{\tilde{f}*}  & 1+\Delta_3^{\tilde{f}}\\
\end{matrix}
\right),
\label{eq:mass_param}
\end{equation}
with $\tilde{f} = \widetilde{Q}_L , \widetilde{u}_R, \widetilde{d}_R,
\widetilde{e}_R, \widetilde{L}_L$. 
In the minimal SU(5) GUT, there are relations among the sfermion
mass matrices at the GUT scale:
\begin{equation}
 m^2_{\tilde Q_L} = V_{QU} (m^2_{{\tilde u}_R})^t V_{QU}^\dagger
  =V_{QE} (m^2_{{\tilde e}_R})^t  
 V_{QE}^\dagger
 {\rm ~and~} 
  m^2_{{\tilde d}_R} = V_{DL}^* (m^2_{{\tilde L}_L})^t V_{DL}^t,
\end{equation}
where $V_{QU}$, $V_{QE}$ and $V_{DL}$ are the GUT ``CKM'' matrices,
which are defined in Sec.~\ref{MSGUT}. In this paper, however, we treat
these five mass matrices independently, without restricted to the above
GUT relation, to clarify each effect on proton decay.

\begin{figure}[t]
\begin{center}
\includegraphics[height=55mm]{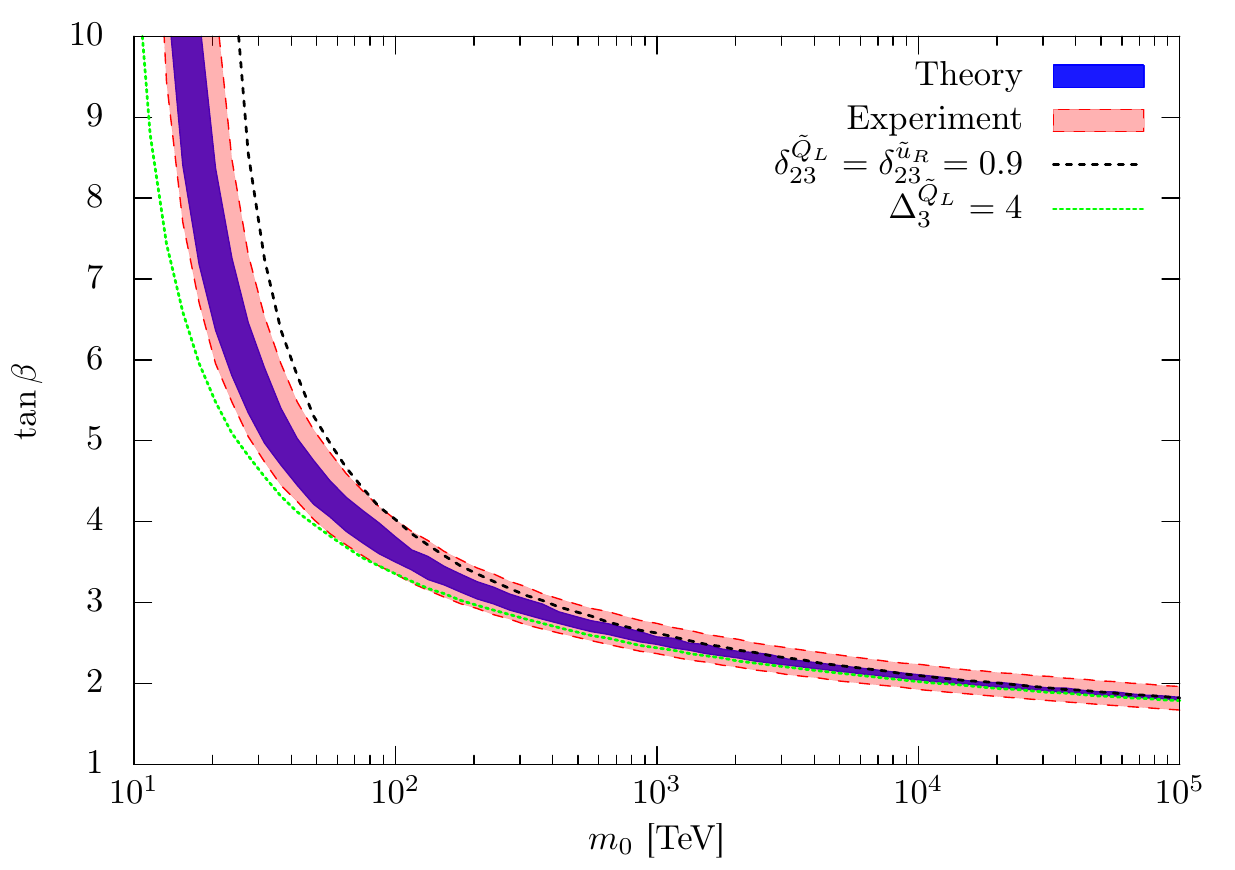}
\caption{$\tan\beta$ as a function of $m_0$ for the observed Higgs mass.
Red and blue bands show the experimental and theoretical uncertainties,
 respectively, for $\mu_H=m_0=m_{A^0}$ and all $\delta$'s and $\Delta$'s
 are set to be zero. The gaugino masses are set to be
 $M_{\tilde{B}}=600$~GeV, $M_{\tilde{W}}=300$~GeV, and
 $M_{\tilde{g}}=-2$~TeV. The cases of $\delta^{\tilde
 Q_L}_{13}=\delta^{\tilde u_R}_{13}=0.9$ (black line) and $
 \Delta^{\tilde Q_L}_{3}=4$ (green line) are also shown.}
\label{fig:tanb}
\end{center}
\end{figure}

As we will see, the proton decay rate has strong dependence on
$\tan\beta$. In Fig.~\ref{fig:tanb}, we show the predicted $\tan\beta$
for the observed Higgs mass as a function of the sfermion mass scale
$m_0$. The red and blue bands show the 
experimental and theoretical uncertainties, respectively, for
$\mu_H=m_0=m_{A^0}$ and all $\delta$'s and $\Delta$'s are zero. For the
experimental inputs, see Table \ref{tab:inputs} in
Appendix~\ref{sec:input}. We estimate the theoretical error by changing
the scale of matching between the MSSM and the (SM+gauginos) system from
$m_0/3$ to $3 m_0$. The gaugino masses are set to be
$M_{\tilde{B}}=600$~GeV, $M_{\tilde{W}}=300$~GeV, and
$M_{\tilde{g}}=-2$~TeV. We also show the cases of $\delta^{\tilde
Q_L}_{13}=\delta^{\tilde u_R}_{13}=0.9$ (black line) and $
\Delta^{\tilde Q_L}_{3}=4$ (green line). In this estimation, we use the
two-loop renormalization group equations (RGEs) in the (SM + gauginos)
system and the one-loop threshold effects from heavy sfermions and
higgsinos. This figure illustrates that a relatively small value of
$\tan \beta$ is favored in the high-scale SUSY scenario.

\subsection{Flavor Constraints}

The soft SUSY-breaking terms in general introduce new sources of flavor
and CP violation, which are severely restricted by low-energy precision
experiments \cite{Gabbiani:1996hi}. As we will see, the flavor violation
of squarks can strongly affect proton decay, and the slepton flavor
violation not so much. In the rest of the section, we briefly review the
current experimental constraints on the squark flavor mixing.

\subsubsection{Meson Mixing}

\begin{figure}[t]
\begin{center}
\includegraphics[width=80mm,clip]{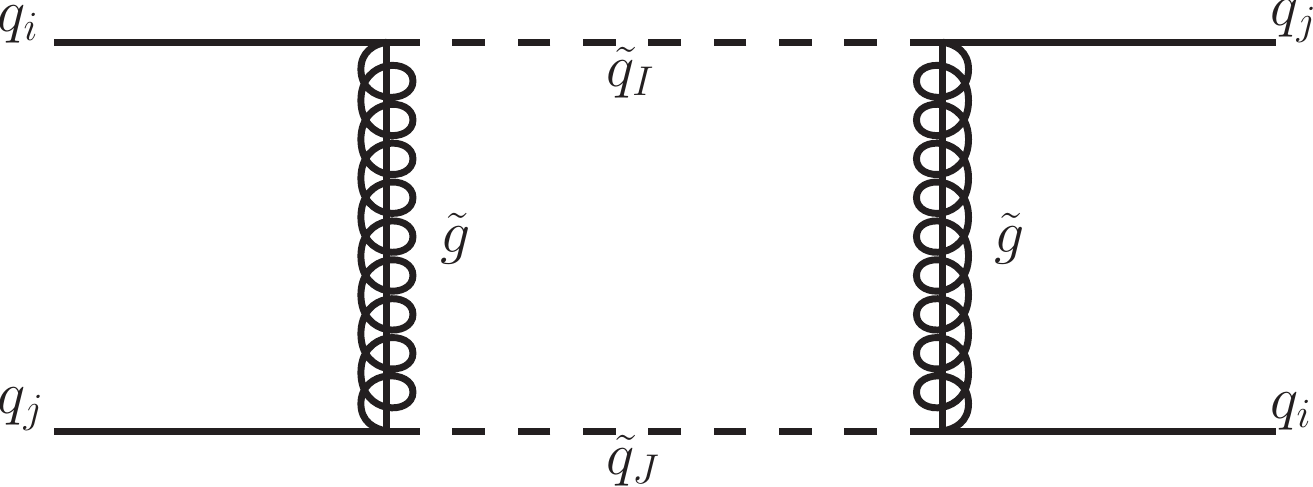}
\caption{An example of the dominant diagram contributing to the meson mixings 
in the presence of the squark flavor mixing. }
\label{fig:box}
\end{center}
\end{figure}

The  $\Delta F=2$ meson mixings give strong constraints on the flavor
violation $\delta$'s. The dominant contribution comes from the box
diagram of Fig.~\ref{fig:box}. The contribution to the oscillation is
represented by the following $\Delta F = 2$ effective Hamiltonian, 
\begin{eqnarray}
 H_{\rm eff} = \sum_{A=1}^5 C_A O_A + \sum_{A=1}^3 \tilde C_A \tilde O_A,
\end{eqnarray}
where the operators $O_A$ and $\tilde O_A$ are defined as follows:
\begin{eqnarray}
 && O_1 = (\bar q^\alpha_{Li} \gamma_\mu q^\alpha_{Lj}) 
 (\bar q^\beta_{Li} \gamma^\mu q^\beta_{Lj}), \nonumber\\
 && O_2 = (\bar q^\alpha_{Ri} q^\alpha_{Lj}) (\bar q^\beta_{Ri} q^\beta_{Lj}),~~~
 O_3 = (\bar q^\alpha_{Ri} q^\beta_{Lj}) (\bar q^\beta_{Ri} q^\alpha_{Lj}), \nonumber\\
 && O_4 = (\bar q^\alpha_{Ri} q^\alpha_{Lj}) (\bar q^\beta_{Li} q^\beta_{Rj}),~~~
 O_5 = (\bar q^\alpha_{Ri} q^\beta_{Lj}) (\bar q^\beta_{Li} q^\alpha_{Rj}),
 \label{eq:FV-operator}
\end{eqnarray}
and $\tilde O_A$ by $R \leftrightarrow L$. In the large squark-mass
limit, $m_{\tilde q} \gg m_{\tilde g}$, the dominant SUSY contributions
to the Wilson coefficients are approximately given by 
\begin{eqnarray}
 && C_1 \simeq \frac{11\alpha_3^2}{36} H(m^2_{\tilde q_{LI}},m^2_{\tilde q_{LJ}}) (R^\dagger_{\tilde{q}_L})_{iI} (R_{\tilde{q}_L})_{Ij} (R^\dagger_{\tilde{q}_L})_{iJ} (R_{\tilde{q}_L})_{Jj},\nonumber
\\
 && C_4 \simeq - \frac{\alpha_3^2}{3}  H(m^2_{\tilde q_{RI}},m^2_{\tilde q_{LJ}}) (R^\dagger_{\tilde{q}_R})_{iI} (R_{\tilde{q}_R})_{Ij} (R^\dagger_{\tilde{q}_L})_{iJ} (R_{\tilde{q}_L})_{Jj}
  , ~~ C_5 \simeq -\frac{5}{3} C_4
,\nonumber \\
 && \tilde C_1 \simeq\frac{11\alpha_3^2}{36} H(m^2_{\tilde q_{RI}},m^2_{\tilde q_{RJ}}) (R^\dagger_{\tilde{q}_R})_{iI} (R_{\tilde{q}_R})_{Ij} (R^\dagger_{\tilde{q}_R})_{iJ} (R_{\tilde{q}_R})_{Jj},
 \end{eqnarray}
where $H(x,y) = \log(x/y)/(x-y)$ and $R$'s are unitary matrices defined
in Eq. (\ref{eq:mixing}). The other Wilson coefficients $C_2, C_3,
\tilde C_2 $ and $\tilde C_3$ are less significant in the present model.

\begin{figure}[t]
\begin{center}
\subfigure[]{\includegraphics[width=75mm]{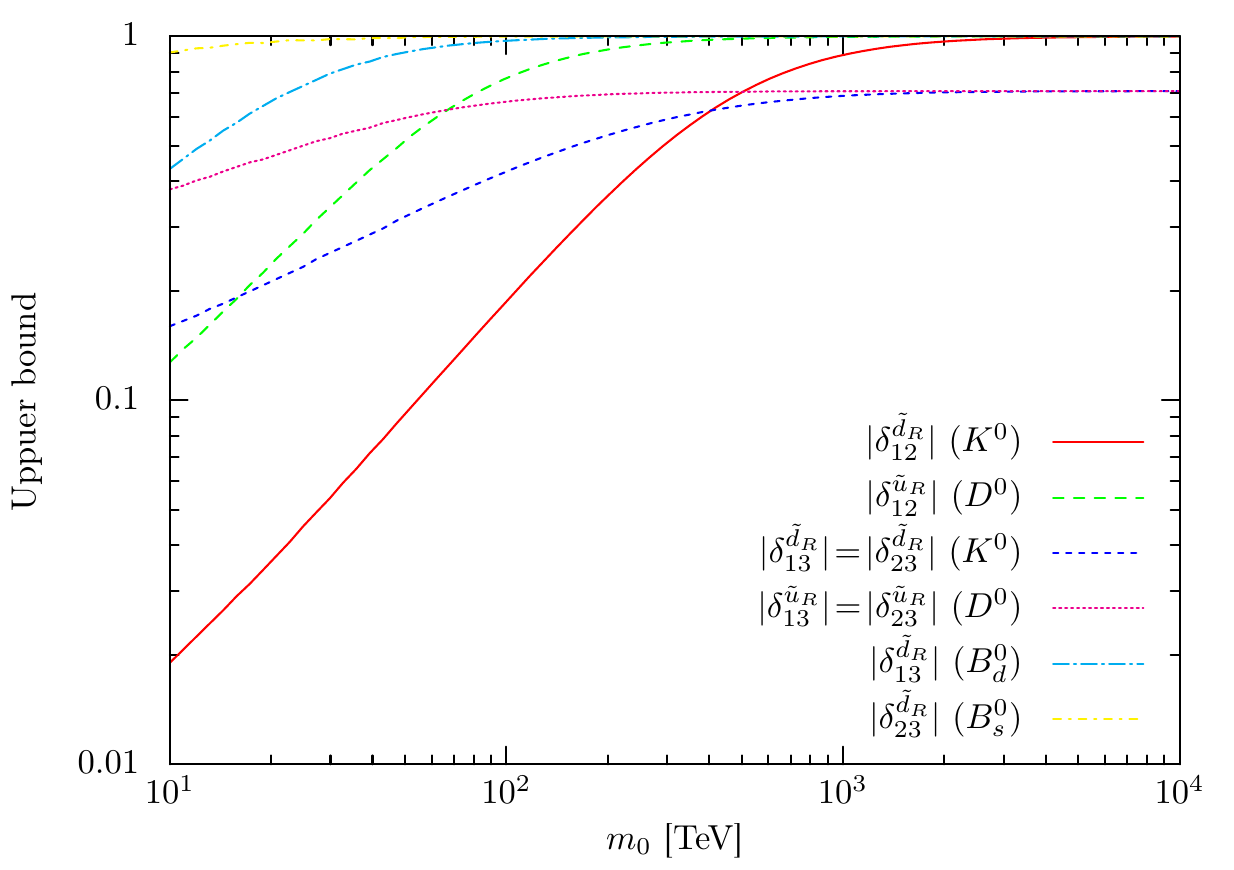}}
\subfigure[]{\includegraphics[width=75mm]{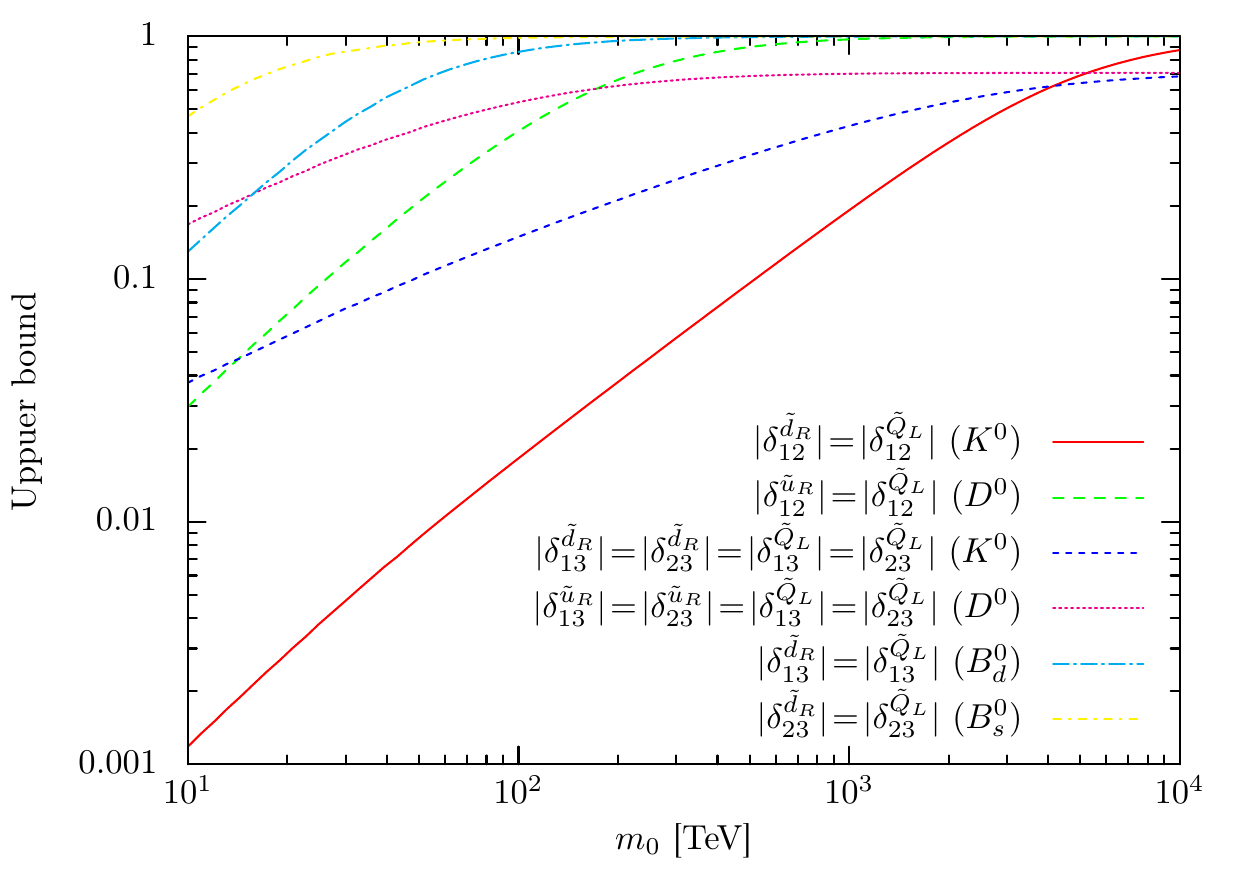}}
\caption{Upper-bound on the flavor violating mass terms $\delta$.
(a): One chirality flavor violation
(b): Both chirality flavor violation.
We choose the ``worst'' case of the CP phases
and take $M_{\tilde B} = 600$~GeV, $M_{\tilde W} = 300$~GeV, and
 $M_{\tilde g} = - 2$~TeV. } 
\label{fig:constraint}
\end{center}
\end{figure}

In Fig.~\ref{fig:constraint}, we show the constraints on $\delta$'s from
the meson mixings. The left (right) panel illustrates the case where
flavor violation occurs in either (both) chirality. To get the
constraints, we evolve the Wilson coefficients down to relevant hadronic
scale and then use the results of new physics fits of
Refs.~\cite{Bona:2007vi,  Bevan:2012waa,UTfit}. The CP phase is chosen
so that the strongest constraint is to be obtained. We set $M_{\tilde B} =
600$~GeV, $M_{\tilde W} = 300$~GeV, and $M_{\tilde g} = - 2$~TeV in this
plot. It is found that especially $\delta^{\widetilde{Q}_L}_{12}$ and
$\delta^{\widetilde{d}_R}_{12}$ are stringently restricted from the
$K^0$-$\bar{K}^0$ mixing even in the case of high-scale SUSY. Other
flavor-violating parameters are allowed to be sizable when
$m_0>10^2$~TeV. In the absence of CP violation, these constraints
get less. Especially, constraints from $K^0$-$\bar K^0$ and $D^0$-$\bar
D^0$ are greatly relaxed in the case of CP conservation, which allows
$\delta$'s to be ${\cal O}(10)$ times larger.

\subsubsection{EDM}
\label{sec:edmconstraint}

\begin{figure}[t]
\begin{center}
\includegraphics[width=80mm,clip]{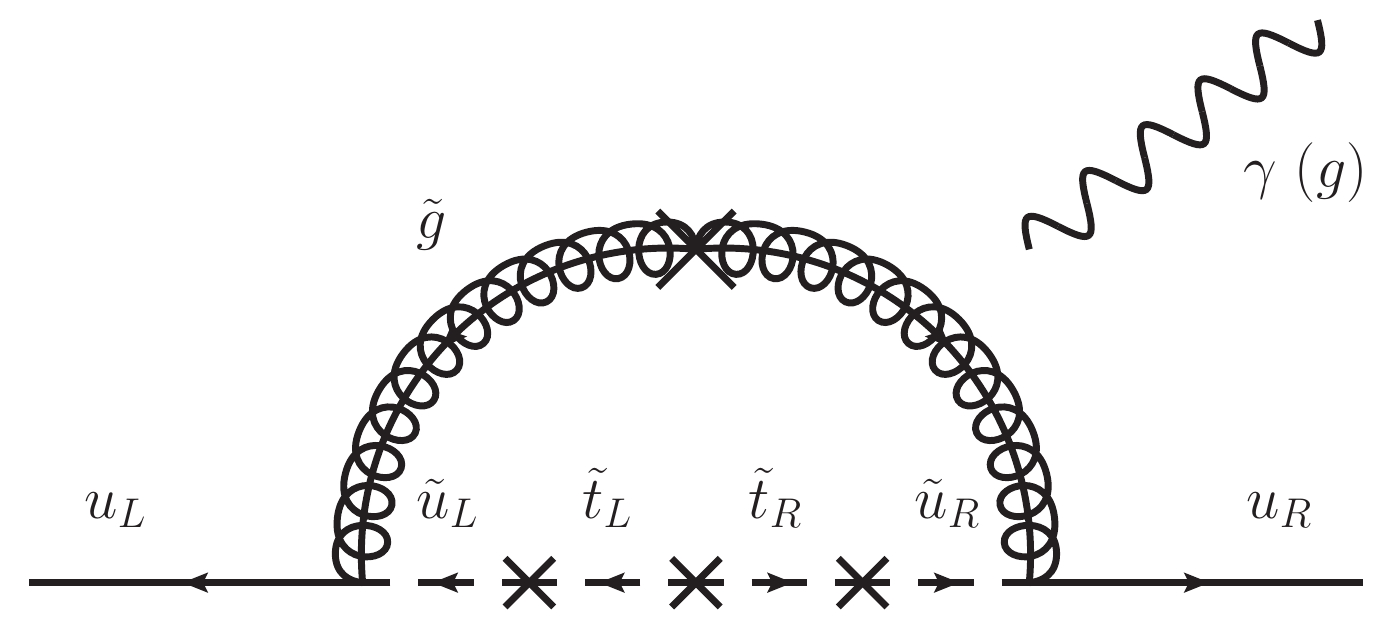}
\caption{An example of the dominant diagram contributing to the EDMs and
 CEDMs of light quarks in the presence of the squark flavor mixing. }
\label{flvCEDM}
\end{center}
\end{figure}

\begin{figure}[t]
\begin{center}
\includegraphics[height=65mm,clip]{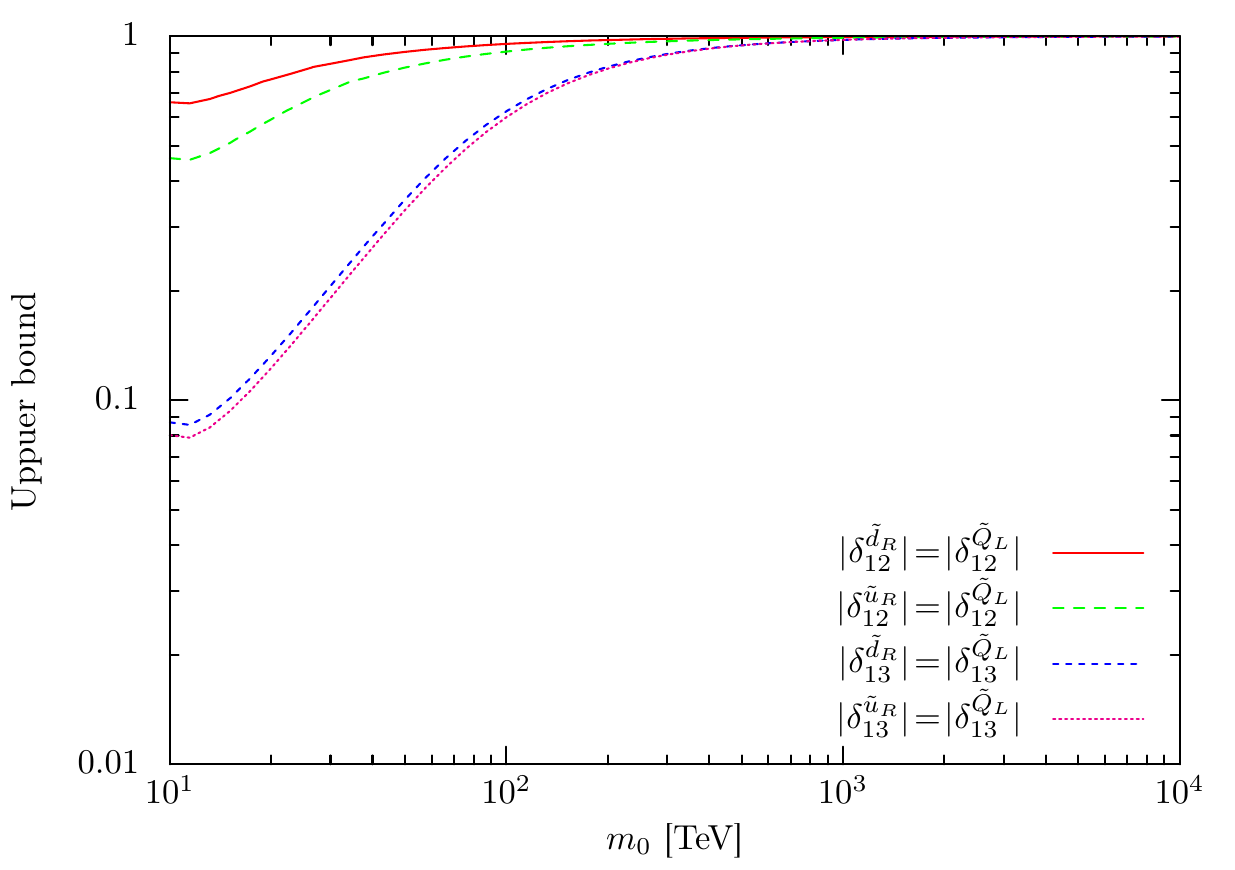}
\caption{Constraints on flavor mixing parameters as functions of
 sfermion mass scale $m_0$. Purple, blue, red, and green lines
 illustrate constraints on $\vert \delta_{13}^{\widetilde{Q}_L}\vert =
 \vert \delta_{13}^{\widetilde{u}_R}\vert$, $\vert
\delta_{13}^{\widetilde{Q}_L}\vert = \vert
\delta_{13}^{\widetilde{d}_R}\vert$, $\vert
\delta_{12}^{\widetilde{Q}_L}\vert = \vert
\delta_{12}^{\widetilde{u}_R}\vert$, and
$\vert \delta_{12}^{\widetilde{Q}_L}\vert = \vert
\delta_{12}^{\widetilde{d}_R}\vert$, respectively.  We take $M_{\tilde
 B} = 600$~GeV, $M_{\tilde W} = 300$~GeV, $M_{\tilde g} = - 2$~TeV, $\tan\beta = 5$ and
 $\mu_H=m_0$. } 
\label{fig:nedm}
\end{center}
\end{figure}

In the presence of CP violation, the electric dipole moments (EDMs)
provide stringent limits on the flavor mixing in the sfermion masses,
though the EDMs are flavor-conserving quantities in nature. As we shall
see below, the dimension-five proton decay rate is quite sensitive to the
squark flavor violation, which is constrained by the neutron
EDM.\footnote{
EDMs of diamagnetic atoms, such as the EDM of mercury, also provide
similar constraints on the squark flavor violation, which are comparable
to those from the neutron EDM within the theoretical uncertainty. 
} On the
assumption of the Peccei-Quinn mechanism \cite{Peccei:1977hh} to solve
the strong CP problem, the relevant effective operators of the lowest mass
dimension are the EDMs
and chromoelectric dipole moments (CEDMs) of light quarks.\footnote{
However, the contribution of the dimension-six Weinberg operator
\cite{Weinberg:1989dx} might be comparable to that of EDMs and CEDMs. In
the present case, however, the operator is induced at ${\cal
O}(\alpha_3^2)$, and thus can be neglected in the leading order
calculation.} The CP violating effects induced by squarks are included
into these two quantities.
In Fig.~\ref{flvCEDM}, we show an example of the diagrams
which yield the EDMs and the CEDMs. As illustrated in the diagram, the dominant
contribution is given by
the flavor-violating processes, where the mass terms of heavy quarks,
especially that of top quark, flip the chirality. For instance, the EDM
$d_u$ and CEDM $\tilde{d}_u$ of up quark are approximately give
as\footnote{These approximate formulae, in particular that for the EDM,
do not work well as squark mass is taken to be larger, though; in such a
case the mixing effect of the CEDM into the EDM becomes dominant
\cite{Degrassi:2005zd, Fuyuto:2013gla}. In our calculation, we include
the effect by using the renormalization group equations.}
\begin{align}
 d_u&\simeq 
 -\frac{4}{3}\frac{\alpha_3}{4\pi}eQ_u\frac{m_t}{m_0^4} 
{\rm Im}\bigl[\mu_H M_{\widetilde{g}} \cot\beta \delta^{\widetilde{Q}_L}_{13}
\delta^{\widetilde{u}_R*}_{13}
\bigr]~,\nonumber \\
\tilde{d}_u&\simeq
6\frac{\alpha_3}{4\pi}\frac{m_t}{m_0^4}
\ln \biggl(\frac{m_0}{\vert M_{\widetilde{g}}\vert}\biggr)
{\rm Im}\bigl[\mu_H M_{\widetilde{g}} \cot\beta \delta^{\widetilde{Q}_L}_{13}
\delta^{\widetilde{u}_R*}_{13}
\bigr]~,
\end{align}
with $eQ_u$ the charge of up quark. Similar expressions hold for down
and strange quarks. Notice that both the left-handed and right-handed
squark mixings are required to utilize the enhancement by heavy quark
masses.  
By evaluating the contribution with the renormalization group improved
method described in Ref.~\cite{Fuyuto:2013gla}, we obtain constraints on
the flavor mixing parameters from the current experimental bound on the
neutron EDM, $\vert d_n\vert < 2.9\times 10^{-26}~e\cdot {\rm cm}$
\cite{Baker:2006ts}. The results are shown in Fig.~\ref{fig:nedm}. In
the figure, the purple, blue, red, and green lines show the constraints
on $\vert \delta_{13}^{\widetilde{Q}_L}\vert = \vert
\delta_{13}^{\widetilde{u}_R}\vert$, $\vert
\delta_{13}^{\widetilde{Q}_L}\vert = \vert
\delta_{13}^{\widetilde{d}_R}\vert$, $\vert
\delta_{12}^{\widetilde{Q}_L}\vert = \vert
\delta_{12}^{\widetilde{u}_R}\vert$, and
$\vert \delta_{12}^{\widetilde{Q}_L}\vert = \vert
\delta_{12}^{\widetilde{d}_R}\vert$, respectively, as functions of the
sfermion mass scale $m_0$. 
We take $M_{\tilde B} = 600$~GeV, $M_{\tilde W} = 300$~GeV, $M_{\tilde g}
= - 2$~TeV, and $\mu_H=m_0$. 
In the calculation, we use
\begin{equation}
 d_n=0.79 d_d-0.20d_u+e(0.30\tilde{d}_u+0.59\tilde{d}_d)~
\label{dnPQ_sum_rule}
\end{equation}
to estimate the neutron EDM, which is obtained by using the method of the QCD
sum rules \cite{Hisano:2012sc}.\footnote{
When one imposes the Peccei-Quinn symmetry, the strange CEDM
contribution to the neutron EDM completely vanishes in the case of the
sum-rule computation. Therefore, $\delta^{\widetilde{Q}_L}_{23}$ and
$\delta^{\widetilde{d}_R}_{23}$ are not constrained. 
This may indicate that the sum-rule calculation
does not include the strange-quark contribution appropriately. In fact,
the contribution is expected to be sizable from the estimation based on
the chiral perturbation theory \cite{Fuyuto:2012yf}. At this moment,
both methods have large uncertainty and no consensus has been reached
yet.} 
The figure illustrates that ${\cal O}(1)$
flavor mixing results in constraints on the sfermion mass scale as high
as ${\cal O}(10^2)$~TeV.

\section{Proton Decay with Sfermion Flavor Violation}
\label{sec:protondecay}

\subsection{Minimal SUSY SU(5) GUT}
\label{MSGUT}

In this section, we give a short review on the minimal SUSY SU(5) GUT
\cite{Dimopoulos:1981zb, Sakai:1981gr} to clarify our notation and
conventions used in this paper. 
Just like the Georgi-Glashow SU(5) model \cite{Georgi:1974sy},
the MSSM matter fields are embedded in a $\bar{\bf 5}\oplus {\bf 10}$
representation; the SU(2)$_L$ singlet down-type quarks $\bar{d}_i$ and
doublet leptons $L_i$ are in the $\bar{\bf 5}$ fields, $\Phi_i$, while
the SU(2)$_L$ singlet up-type quarks, $\bar{u}_i$, doublet quarks,
$Q_i$, and singlet leptons, $\bar{e}_i$, are in the ${\bf 10}$
representations, $\Psi_i$. The MSSM Higgs superfields, $H_u$ and $H_d$,
are incorporated into a pair of {\bf 5} and $\bar{\bf 5}$ superfields
and their SU(5) partners $H_C$ and $\bar{H}_{C}$ are called
the color-triplet Higgs multiplets. The gauge vector multiplets are
embedded into an adjoint vector multiplet. The new gauge fields
introduced to form the adjoint representation are called the $X$-bosons,
and they acquire masses of the order of the GUT scale after the SU(5)
gauge group is broken into the SM gauge group by the VEV of an adjoint
Higgs boson. 

In the minimal SUSY SU(5) GUT, the Yukawa interactions originate from
the following superpotential:
\begin{align}
 W_{\rm Yukawa} &=
\frac{1}{4}h^{ij}\epsilon_{\hat{a}\hat{b}\hat{c}\hat{d}\hat{e}}\Psi_i^{\hat{a}
 \hat{b}} \Psi_j^{\hat{c}\hat{d}}H^{\hat{e}} -\sqrt{2}
f^{ij}\Psi_i^{\hat{a}\hat{b}} \Phi_{j\hat{a}}\bar{H}_{\hat{b}}~,
\label{superpotentialYukawa}
\end{align}
where $\hat{a},\hat{b},\dots=1$--$5$ represent the SU(5) indices;
$\epsilon_{\hat{a}\hat{b}\hat{c}\hat{d}\hat{e}}$ is the totally
antisymmetric tensor with $\epsilon_{12345}=1$; $h^{ij}$ is symmetric
with respect to the generation indices $i,j$. The field re-definition of
$\Psi$ and $\Phi$ reveals that the number of the physical degrees of
freedom in $h^{ij}$ and $f^{ij}$ is twelve. Among
them, six is for quark mass eigenvalues and four is for the CKM matrix
elements, so we have two additional phases \cite{Ellis:1979hy}. 

These Yukawa terms are matched to the MSSM Yukawa couplings at the GUT
scale. Note that the generation basis of the MSSM superfields may be
different from that of the SU(5) superfields $\Psi_i$ and $\Phi_i$. To
take the difference into account, we write the relation between the
SU(5) components and the MSSM superfields as
\begin{align}
 \Psi_i&\ni \{
Q_i,~(V_{QU}^{})_{ij}\overline{u}_j,~(V_{QE}^{})_{ij}\overline{e}_j
\}~,\nonumber \\
\Phi_i&\ni\{
\overline{d}_i,~(V_{DL}^{})_{ij}L_j
\}~,
\end{align}
where $V_{QU}^{}$, $V_{QE}^{}$, and $V_{DL}^{}$ are unitary
matrices, which play a similar role to the CKM matrix. In this paper,
we take them as 
\begin{equation}
 V_{QU}=P^*~,~~~~~~V_{QE}=V_{\rm CKM}(M_{\rm GUT})~,~~~~~~V_{DL}=\1~,
\label{eq:GUT_CKM}
\end{equation}
where $P$ is a diagonal phase matrix with $\det P=1$ and $V_{\rm
CKM}(M_{\rm GUT})$ is the CKM matrix at the GUT scale. 
Then, we have the matching condition as follows:
\begin{align}
& h^{ij}=(P\hat{f}_u(M_{\rm GUT}))^{ij}~,\nonumber \\
&f^{ij}=(V^*\hat{f}_d(M_{\rm GUT}))^{ij}~,\nonumber \\
&\hat{f}_d(M_{\rm GUT}) =\hat{f}_e(M_{\rm GUT})~,
\label{eq:unification}
\end{align}
where $\hat{f}_u$, $\hat{f}_d$, and $\hat{f}_e$ are diagonal and
non-negative Yukawa matrices of the up-type quarks, the down-type
quarks, and the charged leptons, respectively, and
$V \equiv V_{\rm CKM}(M_{\rm GUT})$. In this basis, the Yukawa terms are
written in terms of the MSSM superfields as
\begin{align}
 W_{\rm Yukawa}&=(\hat{f}_u)^{ij}
(Q^{a}_i\cdot H_u)\overline{u}_{ja}-(V^*\hat{f}_d)^{ij} (Q^{a}_i\cdot H_d)
\overline{d}_{ja}-(\hat{f}_e)^{ij} 
\overline{e}_i (L_j\cdot H_d)\nonumber \\
&-\frac{1}{2}(P\hat{f}_u)^{ij}\epsilon_{abc}
(Q^{a}_i \cdot Q^{b}_j) H^c_C
+(V^*\hat{f}_d)^{ij}(Q^{a}_i\cdot L_j)\overline{H}_{Ca}
\nonumber \\
&+(\hat{f}_uV)^{ij}\overline{u}_{ia}
\overline{e}_jH^a_C
-(P^*V^*\hat{f}_d)^{ij}\epsilon^{abc}
\overline{u}_{ia}\overline{d}_{jb}\overline{H}_{Cc}~.
\label{eq:wyukawamssm}
\end{align}
Here, $(A\cdot B)\equiv \epsilon^{\alpha\beta}A_\alpha^{} B_\beta^{}$
with $\alpha, \beta$ representing the SU(2)$_L$ indices, and $a,b,c$
denote the color indices. As it can be seen from the above expression,
we have chosen our basis so that the Yukawa couplings of the up-type
quarks and the charged leptons are diagonalized.

\subsection{Dimension-Five Proton Decay}
\label{sec:dim5formalism}

Now we discuss the proton decay via the color-triplet Higgs
exchange . We first give a set of formulae used in the following
calculation of the proton decay rate.
\begin{figure}[t]
\begin{center}
\includegraphics[height=45mm]{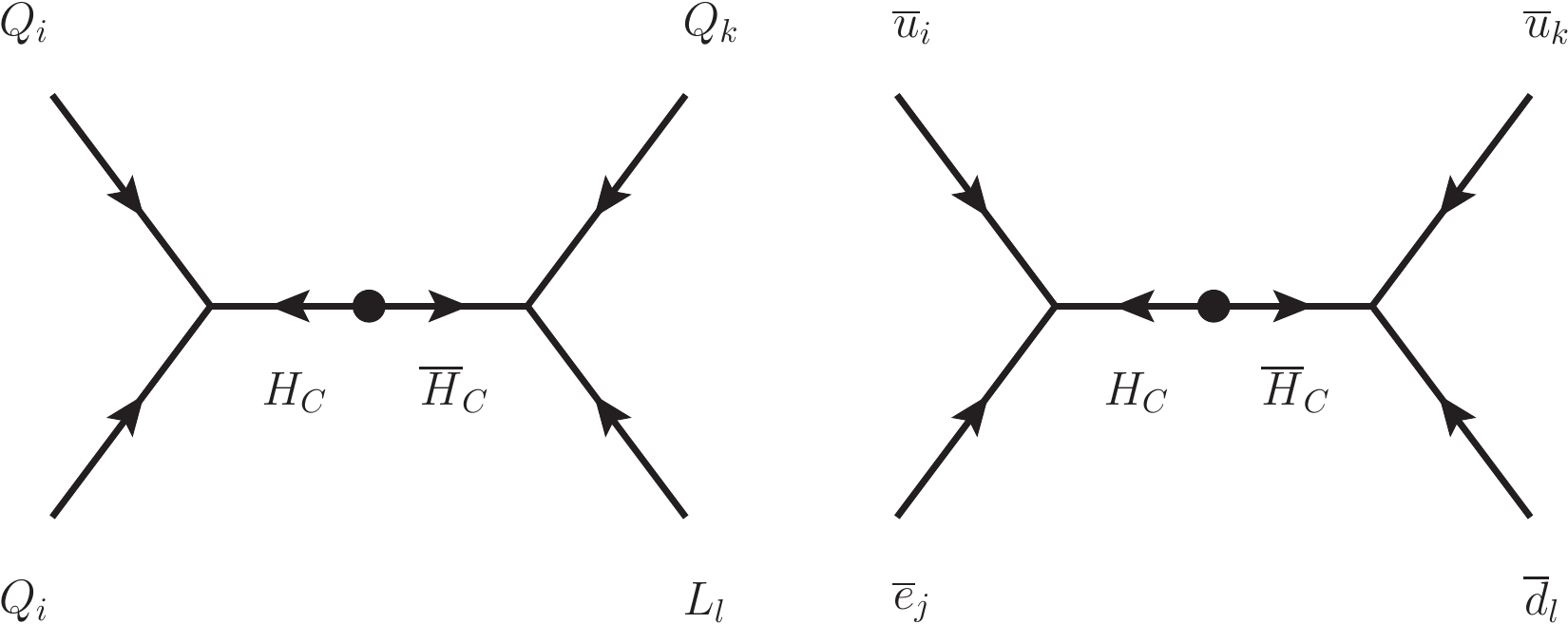}
\caption{Supergraphs for color-triplet Higgs exchanging
  processes where dimension-five effective operators for proton decay
  are induced. Bullets indicate color-triplet Higgs mass term.}
\label{fig:dim5}
\end{center}
\end{figure}
The Yukawa interactions of color-triplet Higgs multiplets, which are
displayed in Eq.~\eqref{eq:wyukawamssm}, give rise to the dimension-five
proton decay operators \cite{Sakai:1981pk,
Weinberg:1981wj}. The diagrams which induce the operators
are illustrated in Fig.~\ref{fig:dim5}. By integrating
out the color-triplet Higgs multiplets, we obtain the effective Lagrangian
\begin{equation}
{\cal L}_5^{\rm eff}= C^{ijkl}_{5L}{\cal O}^{5L}_{ijkl}
+C^{ijkl}_{5R}{\cal O}^{5R}_{ijkl}
~~+~~{\rm h.c.}~,
\end{equation}
where the effective operators ${\cal O}^{5L}_{ijkl}$ and ${\cal
O}^{5R}_{ijkl}$ are defined by
\begin{align}
 {\cal O}^{5L}_{ijkl}&\equiv\int d^2\theta~ \frac{1}{2}\epsilon_{abc}
(Q^a_i\cdot Q^b_j)(Q_k^c\cdot L_l)~,\nonumber \\
{\cal O}^{5R}_{ijkl}&\equiv\int d^2\theta~
\epsilon^{abc}\overline{u}_{ia}\overline{e}_j\overline{u}_{kb}
\overline{d}_{lc}~,
\end{align}
and the Wilson coefficients $C^{ijkl}_{5L}$ and $C^{ijkl}_{5R}$ are
given by
\begin{align}
 C^{ijkl}_{5L}(M_{\rm GUT})&
=+\frac{1}{M_{H_C}}(P\hat{f}_u)^{ij}(V^*\hat{f}_d)^{kl}
~,\nonumber \\
C^{ijkl}_{5R}(M_{\rm GUT})
&=+\frac{1}{M_{H_C}}(\hat{f}_uV)^{ij}(P^*V^*\hat{f}_d)^{kl}
~.
\label{eq:wilson5}
\end{align}
Here, $M_{H_C}$ is the mass of color-triplet Higgs multiplets.
Note that because of the totally antisymmetric tensor in the operators
${\cal O}^{5L}_{ijkl}$ and ${\cal O}^{5R}_{ijkl}$ they must include at
least two generations of quarks. For this reason, the dominant mode of
proton decay induced by the operators is accompanied by strange quarks;
like the $p\to K^+\bar{\nu}$ mode.
The Wilson coefficients in Eq.~\eqref{eq:wilson5} are determined at the
GUT scale. To evaluate the proton decay rate, we need to evolve them
down to low-energy regions by using the RGEs. The RGEs for the
coefficients are presented in Appendix~\ref{sec:rge}.


\begin{figure}[t]
\begin{center}
\includegraphics[height=40mm,clip]{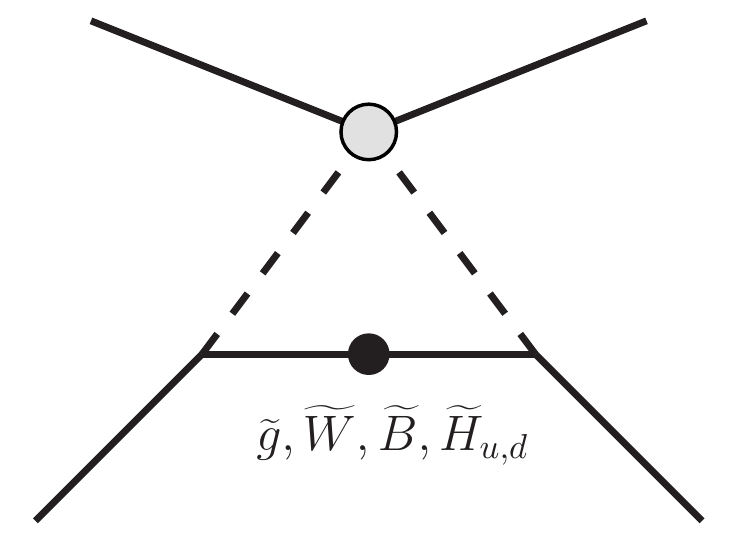}
\caption{One-loop diagram which yields proton decay four-Fermi
 operators. The gray dot indicates the dimension-five effective interactions
 and black dot represents the mass term of exchanged particles; gauginos
 or higgsinos.} 
\label{fig:dim51loop}
\end{center}
\end{figure}

The dimension-five operators contain sfermions in their external lines. 
At the sfermion mass scale $m_0$, sfermions decouple from the
theory, and the dimension-five operators reduce to the dimension-six
four-Fermi operators via the exchange of gauginos and higgsinos. In
Fig.~\ref{fig:dim51loop}, an one-loop diagram which yields the
four-Fermi operators is illustrated. Here, the gray dot indicates the
dimension-five effective interactions and the black dot represents the
mass term of exchanged particles. 
The four-Fermi operators induced here are written in an invariant form
under the SU(3)$_C\otimes$SU(2)$_L\otimes$U(1)$_Y$ symmetry. A set of
such operators is summarized in Refs~\cite{Weinberg:1979sa,
Wilczek:1979hc, Abbott:1980zj}\footnote{
We have slightly changed the labels of the operators as well as the
order of fermions from those presented in Ref.~\cite{Abbott:1980zj}. 
} as follows: 
\begin{align}
 {\cal O}^{(1)}_{ijkl}&=
\epsilon_{abc}(u^a_{Ri}d^b_{Rj})(Q_{Lk}^c\cdot L_{Ll}^{})~,\nonumber \\
 {\cal O}^{(2)}_{ijkl}&=
\epsilon_{abc}(Q^a_{Li}\cdot Q^b_{Lj})(u_{Rk}^ce_{Rl}^{})~,\nonumber \\
{\cal O}^{(3)}_{ijkl}&=
\epsilon_{abc}\epsilon^{\alpha\beta}\epsilon^{\gamma\delta}
(Q^a_{Li\alpha} Q^b_{Lj\gamma})(Q_{Lk\delta}^c L_{Ll\beta}^{})~,\nonumber \\
 {\cal O}^{(4)}_{ijkl}&=
\epsilon_{abc}(u^a_{Ri}d^b_{Rj})(u_{Rk}^c e_{Rl}^{})~.
\label{eq:fourfermidef}
\end{align}
Here we explicitly write the way of contracting the SU(2)$_L$ indices
for ${\cal O}^{(3)}_{ijkl}$. Let us express their Wilson coefficients by
$C^{ijkl}_{(I)}$ for ${\cal O}^{(I)}_{ijkl}$ $(I=1,2,3,4)$. Then, they
are matched with $C_{5L}^{ijkl}$ and $C_{5R}^{ijkl}$ at
the SUSY breaking scale. The matching conditions are summarized in
Appendix~\ref{sec:matchingsusy}. 
Again, the coefficients are evolved down to the electroweak scale
according to the RGEs. The RGEs below the SUSY breaking scale are also
given in Appendix~\ref{sec:rge}. 

Below the electroweak scale $\mu =m_Z$, the effective operators are no
longer invariant under the SU(3)$_C\otimes$SU(2)$_L\otimes$U(1)$_Y$
symmetry; instead, they must respect the SU(3)$_C\otimes$U(1)$_{\rm
em}$, and all of the fields in the operators are to be written in the
mass basis. As mentioned above, the dominant mode of proton decay
induced by the dimension-five effective operators is the $p\to
K^+\bar{\nu}$ mode. The effective Lagrangian which yields the decay mode
is written down as follows:
\begin{align}
 {\cal L}(p\to K^+\bar{\nu}_i^{})
=&C_{RL}(dsu\nu_i)\bigl[\epsilon_{abc}(d_R^as_R^b)(u_L^c\nu_i^{})\bigr]
+C_{RL}(usd\nu_i)\bigl[\epsilon_{abc}(u_R^as_R^b)(d_L^c\nu_i^{})\bigr]
\nonumber \\
+&C_{RL}(uds\nu_i)\bigl[\epsilon_{abc}(u_R^ad_R^b)(s_L^c\nu_i^{})\bigr]
+C_{LL}(dsu\nu_i)\bigl[\epsilon_{abc}(d_L^as_L^b)(u_L^c\nu_i^{})\bigr]
\nonumber \\
+&C_{LL}(usd\nu_i)\bigl[\epsilon_{abc}(u_L^as_L^b)(d_L^c\nu_i^{})\bigr]
+C_{LL}(uds\nu_i)\bigl[\epsilon_{abc}(u_L^ad_L^b)(s_L^c\nu_i^{})\bigr]
~.
\label{efflagmz}
\end{align}
Here, all of the fermions are written in terms of the mass
eigenstates. The matching condition for the Wilson coefficients
$C_{RL}$ and $C_{LL}$ at the electroweak scale are listed in
Appendix~\ref{sec:matchingew}.

The Wilson coefficients are taken down to the hadronic scale $\mu=
2$~GeV, where the matrix elements of the effective operators are
evaluated. The RGEs for the step are given in Appendix \ref{sec:rge}. 
For the hadron matrix elements of the effective operators, we use the results
presented by the lattice QCD calculation \cite{Aoki:2013yxa}. Their
values are listed in Table~\ref{tab:matrixelements} in
Appendix~\ref{sec:input}. 
By using the results, we can
eventually obtain the partial decay width of the $p\to K^+ \bar{\nu}_i$
mode as
\begin{equation}
 \Gamma(p\to K^+\bar{\nu}_i)
=\frac{m_p}{32\pi}\biggl(1-\frac{m_K^2}{m_p^2}\biggr)^2
\vert {\cal A}(p\to K^+\bar{\nu}_i)\vert^2~,
\end{equation}
where $m_p$ and $m_K$ are the masses of proton and kaon,
respectively. The amplitude ${\cal A}(p\to K^+\bar{\nu}_i)$ is given by
the sum of the Wilson coefficients at $\mu=2$~GeV multiplied by the
corresponding hadron matrix elements.

By following a similar procedure, we can also evaluate the partial decay
rates for other modes. The resultant expressions are presented in
Appendix~\ref{sec:othermode}.

\subsection{Results}


\begin{figure}[t]
\begin{center}
\includegraphics[height=60mm]{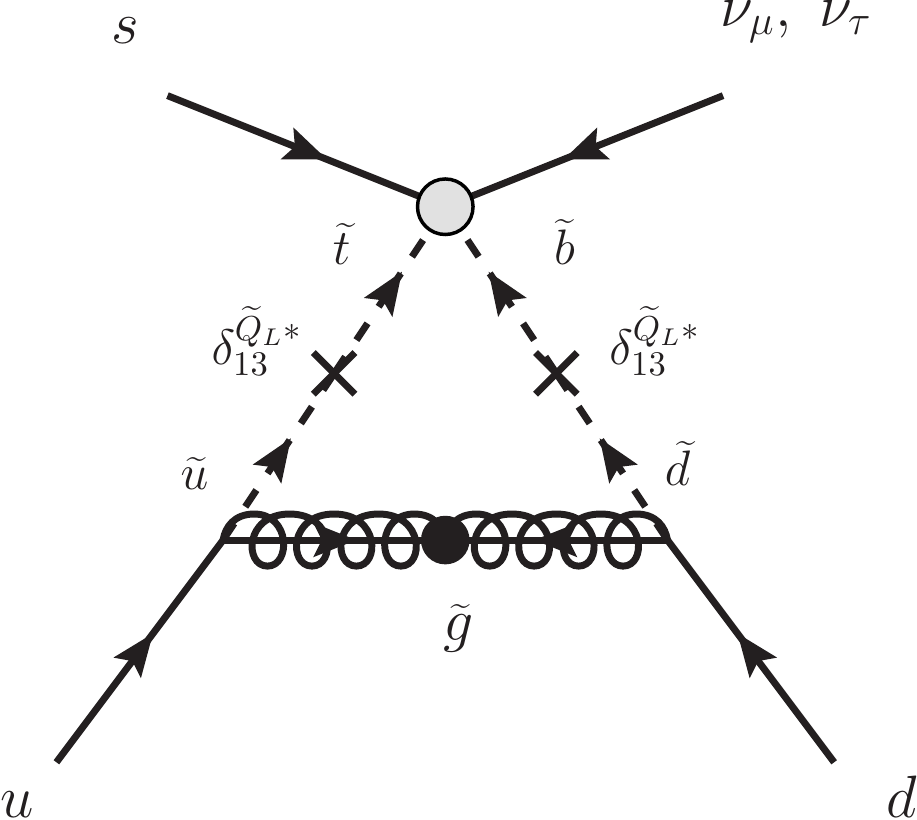}
\caption{Diagram which induces the dominant contribution in the presence
 of the $\delta^{\widetilde{Q}_L}_{13}$ flavor mixing, which is denoted
 by $\times$-mark. }
\label{fig:1loopflvdim5}
\end{center}
\end{figure}

As discussed in Ref.~\cite{Goto:1998qg}, the charged wino and higgsino
exchange processes give rise to the dominant contribution to the
dimension-five proton decay in the case of the minimal flavor violation. 
When the sfermion sector contains sizable flavor violation, on the other
hand, not only the charged fermions, but also the neutral gauginos and
higgsinos can contribute.
Especially, the gluino contribution
becomes significant because of the large value of $\alpha_3$. Since
only the $C^{ijkl}_{(3)}\vert_{\widetilde{g}}$ in
Eq.~\eqref{eq:dim5gluinocontr} contributes to the $p\to K^+\bar{\nu}$
proton decay, the flavor mixing in the mass matrix of
$\widetilde{Q}_L$ is most important; in particular
$\delta^{\widetilde{Q}_L}_{13}$ gives rise to the biggest effects. Let
us estimate the significance. The dominant contribution to the $p\to
K^+\bar{\nu}$ mode is induced by the diagram in
Fig.~\ref{fig:1loopflvdim5}. Here, the cross-mark indicates the flavor
mixing. When the flavor violation is small but sizable, {\it e.g.},
$\delta^{\widetilde{Q}_L}_{13}\sim 0.1$, the contribution is evaluated
as
\begin{align}
 C_{LL}(uds\nu_\mu)&\simeq
-\frac{4}{3}\frac{\alpha_2\alpha_3}{\sin 2\beta}
\frac{m_tm_s}{M_{H_C}m_W^2}\frac{M_{\widetilde{g}}}{m_0^2}~
e^{i\varphi_3}(V_{ud}^{}V_{cs}^{}V_{cs}^*)\bigl(
\delta^{\widetilde{Q}_L*}_{13}\bigr)^2~, \nonumber \\
 C_{LL}(uds\nu_\tau)&\simeq
-\frac{4}{3}\frac{\alpha_2\alpha_3}{\sin 2\beta}
\frac{m_tm_b}{M_{H_C}m_W^2}\frac{M_{\widetilde{g}}}{m_0^2}~
e^{i\varphi_3}(V_{ud}^{}V_{cs}^{}V_{cb}^*)\bigl(
\delta^{\widetilde{Q}_L*}_{13}\bigr)^2~,
\label{eq:approximateglu}
\end{align}
and other Wilson coefficients are found to be sub-dominant. 
Here, we assume $M_{\widetilde{g}}\ll m_0$.
As we have mentioned above, the contribution strongly depends on $\tan\beta$. 
By comparing the results to the higgsino contribution
in the minimal flavor violation case, which is found to be dominant when
$\mu_H \simeq m_0$ \cite{Hisano:2013exa},
we can see that the gluino
contribution becomes dominant when
\begin{equation}
 \bigl\vert \delta^{\widetilde{Q}_L}_{13}\bigr\vert~
\gtrsim~ 2\times 10^{-3}\times
\biggl(\frac{1}{\sin
2\beta}\biggl\vert\frac{\mu_H}{M_{\widetilde{g}}}\biggr\vert\biggr) 
^{\frac{1}{2}}~.
\end{equation}

Before showing the results for the full computation, we briefly comment
on the features of other contributions. The wino and bino contributions
are in general suppressed by the relatively small gauge couplings
compared with the gluino contribution. The higgsino contribution has
already exploited the flavor changing in the Yukawa couplings to make
the most of the enhancement from the third generation Yukawa couplings.
Therefore, the flavor mixing in sfermion masses does not increase the
contribution any more. 

As we will see below, the effects of the other mixing parameters are
generally sub-dominant. In particular, when the flavor violation occurs
only in the slepton sector, the proton decay rate is rarely
changed. This is because the gluino exchange process does not contribute
to the proton decay in such a case. In addition, when only the
right-handed squarks feel the flavor violation, the $p\to K^+\bar{\nu}$
mode is not enhanced because of the same reason. In such a case, on the
other hand, the decay modes including a charged lepton in their final
states, such as the $p\to \pi^0\mu^+$ mode, are considerably
enhanced. We will discuss the feature in more detail below.

\begin{figure}[tp]
\begin{center}
\subfigure[$p\to K^+ \bar{\nu}$]{\includegraphics[width=75mm]{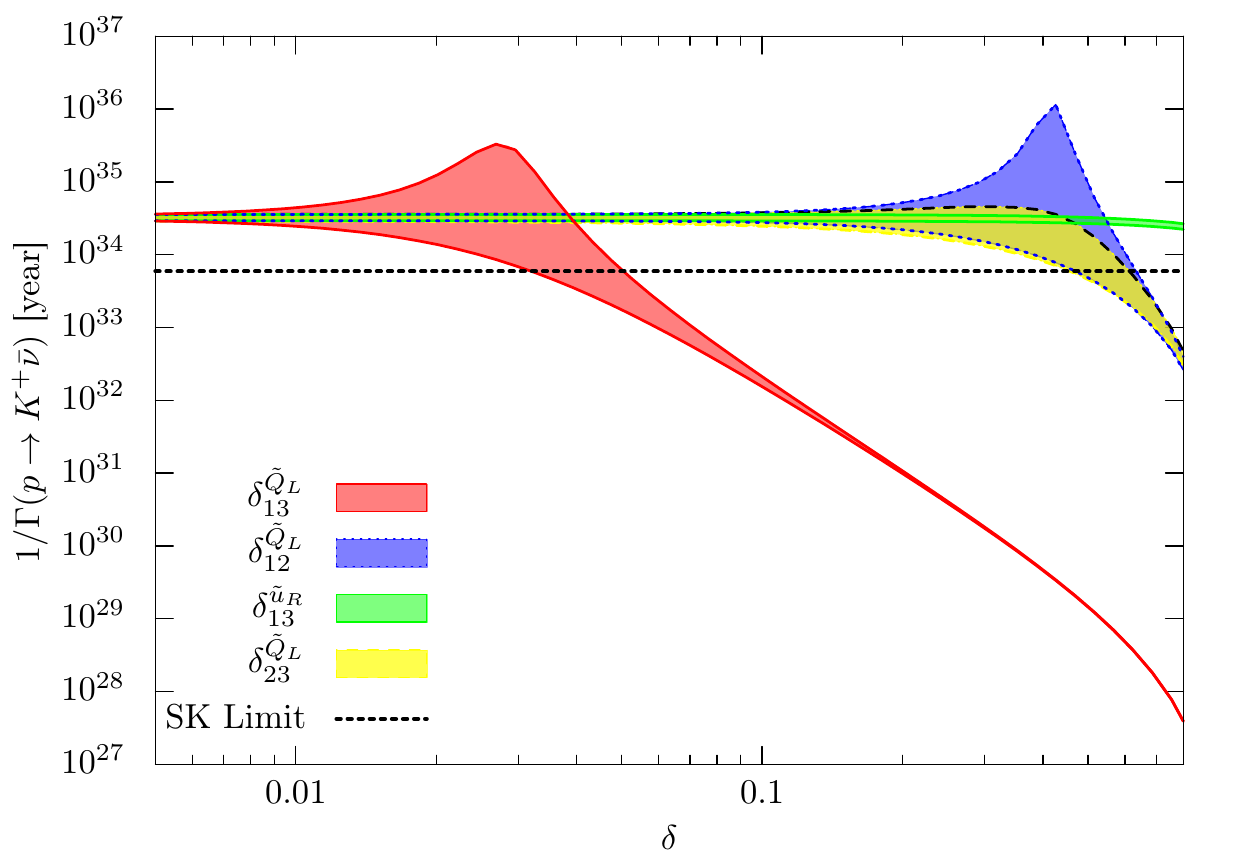}}
\subfigure[$p\to \pi^0 e^+  $]{\includegraphics[width=75mm]{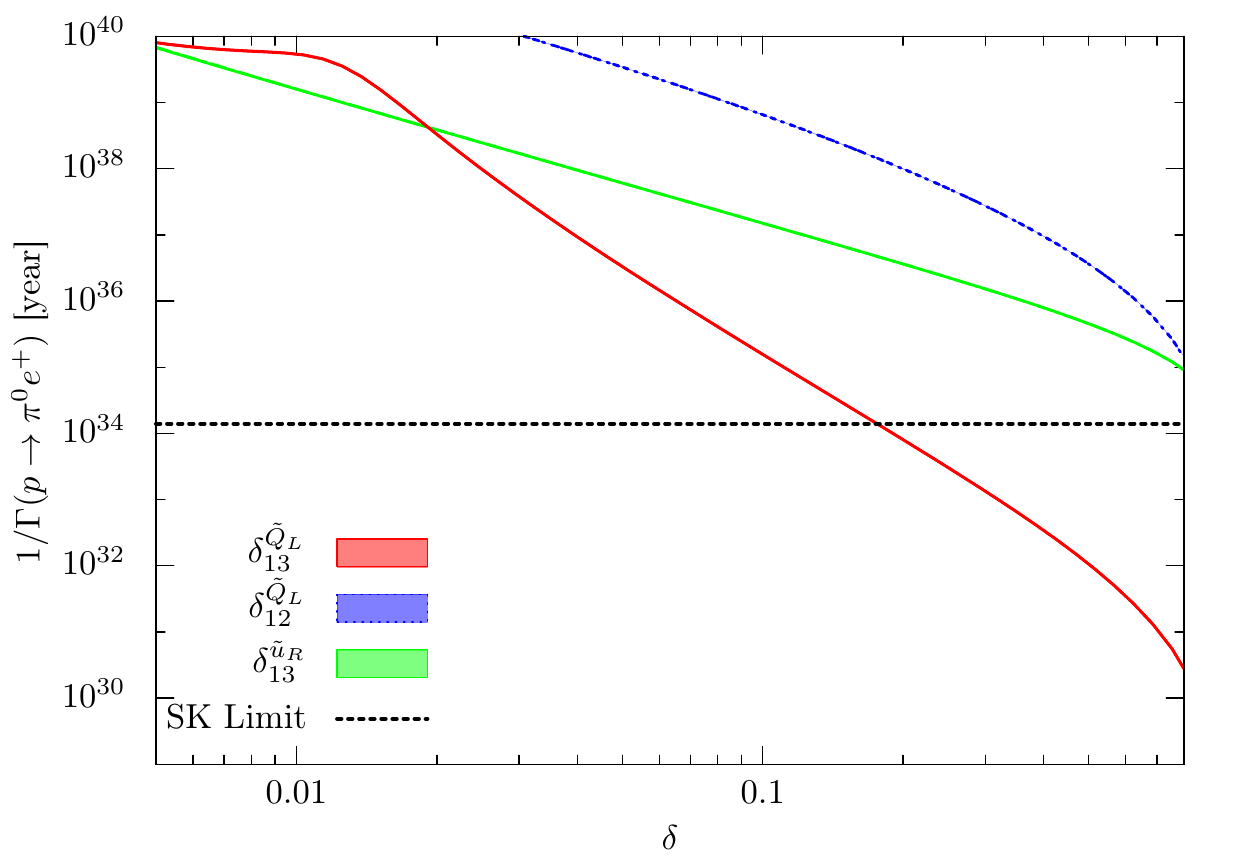}}
\subfigure[$p\to K^0 \mu^+ $]{\includegraphics[width=75mm]{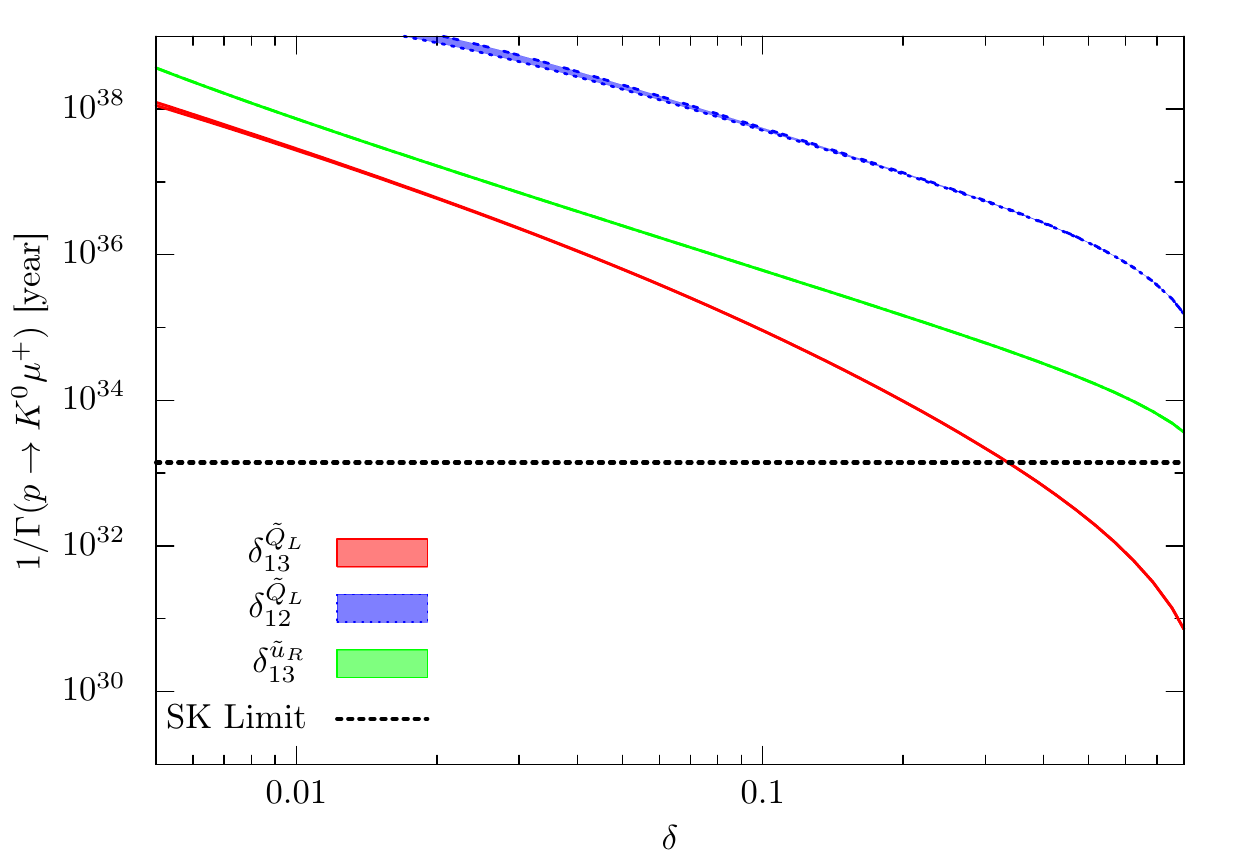}}
\subfigure[$p\to \pi^0 \mu^+ $]{\includegraphics[width=75mm]{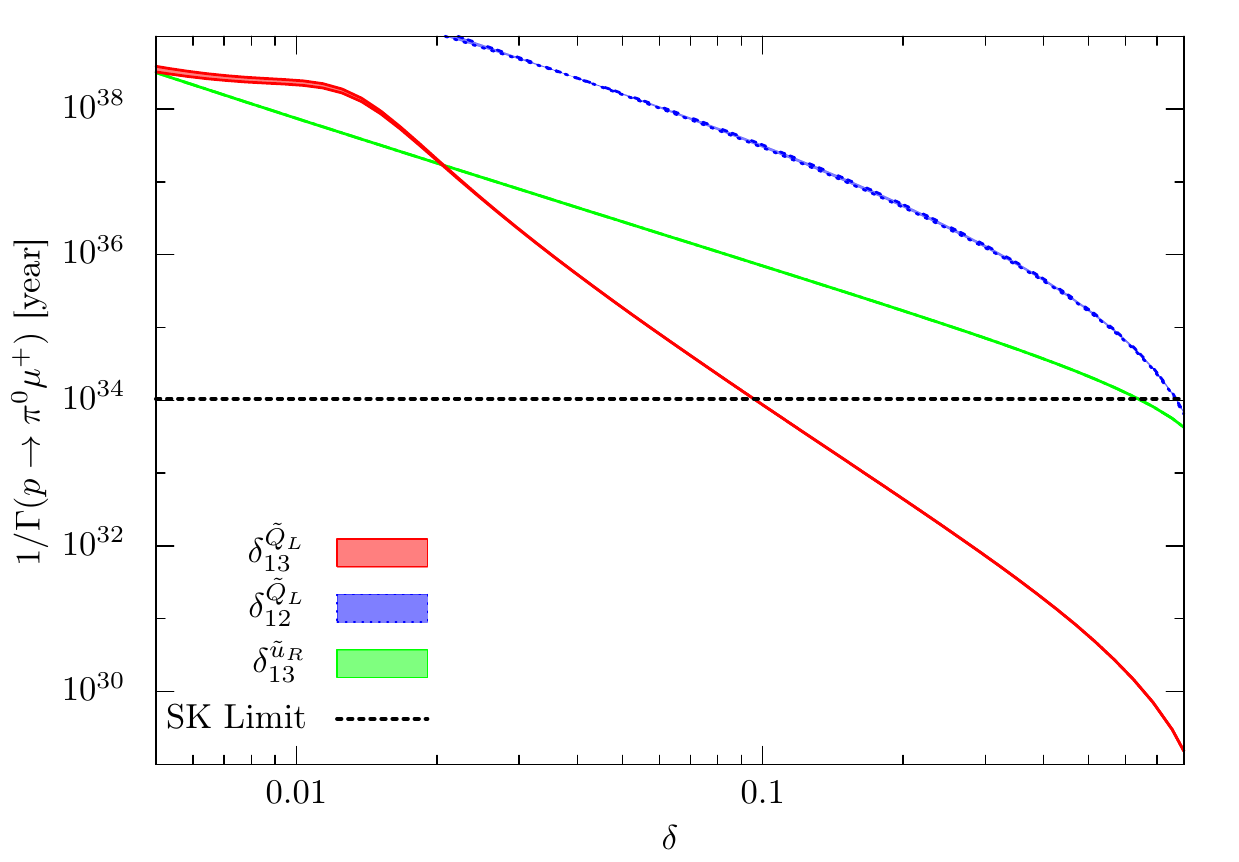}}
\caption{Proton lifetime as functions of flavor mixing parameters
 $\delta$'s. Red, blue, green, and
yellow lines correspond to $\delta^{\widetilde{Q}_L}_{13}$,
$\delta^{\widetilde{Q}_L}_{12}$, $\delta^{\widetilde{u}_R}_{13}$, and
$\delta^{\widetilde{Q}_L}_{23}$, respectively.
The color bands show the uncertainty from unknown CP phases $P$ in the GUT
 Yukawa couplings defined in Eq.~\eqref{eq:GUT_CKM}. We set $m_0=
 100$~TeV, $M_{\tilde B} = 600$~GeV, 
 $M_{\tilde W} = 300$~GeV, $M_{\tilde g} = - 2$~TeV, $\tan\beta=5$, $\mu_H
 = +m_0$, and $M_{H_C} = 10^{16}$~GeV. $\Delta$'s and $\delta$'s which
 are not displayed in the figure are set to be zero. Black dashed lines
 represent the experimental limits presented by Super-Kamiokande
 \cite{Shiozawa,Babu:2013jba}. 
}
\label{fig:life_light}
\end{center}
\end{figure}

Now we show the results. In Fig.~\ref{fig:life_light}, we show the
proton lifetime as functions of selected flavor violating parameters
$\delta$'s in Eq.~(\ref{eq:mass_param}). The red, blue, green, and
yellow lines correspond to $\delta^{\widetilde{Q}_L}_{13}$,
$\delta^{\widetilde{Q}_L}_{12}$, $\delta^{\widetilde{u}_R}_{13}$, and
$\delta^{\widetilde{Q}_L}_{23}$, respectively.
In this figure, the uncertainty
coming from the unknown phases $P$ in the GUT Yukawa couplings defined
in Eq.~\eqref{eq:GUT_CKM} is shown
as color bands. We take $m_0= 100$~TeV, $M_{\tilde B} = 600$~GeV,
$M_{\tilde W} = 300$~GeV , $M_{\tilde g} = - 2$~TeV, and $\tan\beta=5$,
$\mu_H= +m_0$, and $M_{H_C} = 10^{16}$~GeV,\footnote{The color-triplet
Higgs mass $M_{H_C}$ can be as heavy as the GUT scale in the case of the
high-scale SUSY scenario \cite{Hisano:2013cqa}. } and we do not include the
running effects on the gaugino masses. The black dashed lines represent
the experimental limits presented by Super-Kamiokande
\cite{Shiozawa,Babu:2013jba}. From the figure, it is found that
$\delta^{\tilde Q_L}_{13}$ gives strong impacts on the proton lifetime
for each decay channel. It results from the large contribution of
gluino exchange processes to the proton decay rates.
For instance, in the case of the $p \to K^+\bar{\nu}$
decay mode given in the plot (a), gluino dressing parts become dominant
when $\delta^{\tilde 
Q_L}_{13} \gg 0.01$. In the region, the proton partial decay rate is
approximately proportional to the fourth power of $\delta^{\tilde
Q_L}_{13}$, as described in Eq.~\eqref{eq:approximateglu}.
For small $\delta^{\tilde Q_L}_{13} \ll 0.01$, on the other hand,
the higgsino dressing contribution dominates the decay amplitude, and
thus the lifetime hardly depends on the flavor violation.
When $\delta^{\tilde Q_L}_{13} \sim 0.01$, both gluino and higgsino
dressing contributions are comparable to each other, which may result in
a significant cancellation between them, depending on the GUT CP phases
$P$.

We also present the results for the $p\to \pi^0 e^+$, $p\to K^0  \mu^+$,
and $p\to  \pi^0 \mu^+$ channel in the plots (b), (c), and (d) in
Fig.~\ref{fig:life_light}, respectively.
A characteristic feature in this case is
that the right-handed squark flavor violation, such as
$\delta^{\widetilde{u}_R}_{13}$ and $\delta^{\widetilde{d}_R}_{13}$, is
also important. This is because when the final state of proton
decay includes a charged lepton, not
only the operators ${\cal O}^{(1)}_{ijkl}$ and ${\cal O}^{(3)}_{ijkl}$
but also ${\cal O}^{(2)}_{ijkl}$ and ${\cal O}^{(4)}_{ijkl}$ can
contribute to the decay rate. Notice that in the gluino exchange
process the right-handed squark flavor violation can only contribute to
the operator ${\cal O}^{(4)}_{ijkl}$, as can been seen from the formulae
presented in Appendix~\ref{sec:othermode}. For this reason,
$\delta^{\widetilde{u}_R}_{13}$ and $\delta^{\widetilde{d}_R}_{13}$
scarcely affect the anti-neutrino decay modes such as $p\to K^+
\bar{\nu}$, which are induced by the operator ${\cal O}^{(3)}_{ijkl}$,
while they can enhance the charged lepton modes through ${\cal
O}^{(4)}_{ijkl}$. 

\begin{figure}[tp]
\begin{center}
\includegraphics[width=120mm]{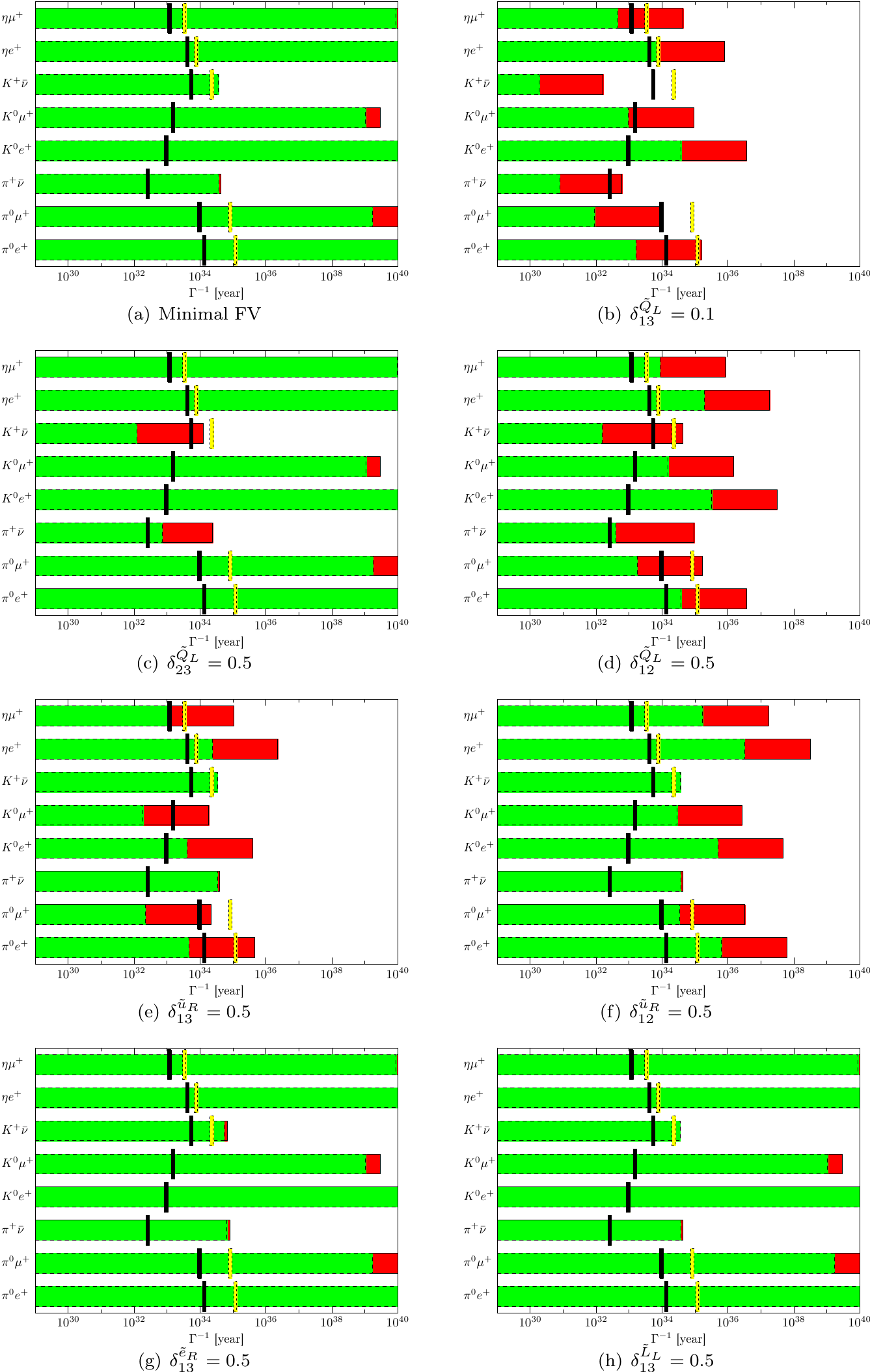}
\caption{
Dependence of the proton decay modes on the flavor structure.
Red bars show the case where a similar set of parameters to those in
 Fig.~\ref{fig:life_light} is taken, while green bars correspond to the
 case in which the gaugino masses are ten times as large as the previous
 ones. Black lines represent the Super-Kamiokande constraints at 90 \%
 CL while yellow lines show the future prospects of
 Hyper-Kamiokande \cite{Shiozawa,Babu:2013jba}.  
}
\label{fig:decay_mode}
\end{center}
\end{figure}

The sfermion flavor violation also alters the branching ratio. This can
be again seen from the plots (b--d) in Fig.~\ref{fig:life_light}; without
flavor violation, the decay rates of these modes are extremely small
compared with that of $p\to K^+\bar{\nu}$, while they become significant
in the presence of sizable flavor violation. To see the feature more clearly,
we show the partial decay rates of selected proton decay modes for various
$\delta$'s in Fig.~\ref{fig:decay_mode}. The red bars show the case in
which we take $m_0= 100$~TeV, $M_{\tilde B} = 600$~GeV, $M_{\tilde W} =
300$~GeV, $M_{\tilde g} = - 2$~TeV, $\tan\beta=5$, $\mu_H= +m_0$, and
$M_{H_C} = 10^{16}$~GeV, while the green bars correspond to the case where the
gaugino masses are ten times as large as the previous ones:
$M_{\tilde B} = 6$~TeV,  $M_{\tilde W} = 3$~TeV, and $M_{\tilde g} = -
20$~TeV. The bar charts in Fig.~\ref{fig:decay_mode} illustrate the
features of the dimension-five proton decay discussed above; 
in the case of the minimal flavor violation, the most significant decay
mode is the $p\to K^+\bar{\nu}$ channel, while other decay modes get also
viable once you switch on the flavor violation;
$\delta^{\widetilde{Q}_L}_{13}$ yields the most significant effects on
the proton decay rate, contrary to the flavor violation in slepton mass
matrices, which gives little contribution; $\delta^{\tilde u_R}_{13}$ enhances
the decay rates of the charged lepton modes, rather than those of the
anti-neutrino modes such as $p\to K^+\bar{\nu}$.


Now let us look for a specific signature of the proton
decay associated with sfermion flavor violation. As one can see from
Fig.~\ref{fig:decay_mode}, in the minimal flavor violation case, only the
anti-neutrino decay modes, $p\to K^+\bar{\nu}$ and $p\to
\pi^+\bar{\nu}$, have sizable decay rates. To distinguish the flavor
violating contribution from it, therefore, we should focus on the
charged lepton decay modes. As shown in Sec.~\ref{sec:dim6xboson},
charged leptonic decay is also induced via the $X$-boson exchanging
process. Since the process is induced by the gauge interactions, the CKM
matrix is the only source for the flavor violation. Thus, in the
$X$-boson exchange contribution, the decay modes which include different
generations in their final states, such as $p\to \pi^0 \mu^+$ and $p\to
K^0e^+$, suffer from the CKM suppression. We will see this feature in
Sec.~\ref{sec:dim6xboson}. Hence, such decay modes can be regarded as
characteristic of extra flavor violation if they are actually
observed. Among them, the experimental
constraint on the $p\to \pi^0 \mu^+$ mode is the severest, and thus it
may offer a good prove for the sfermion flavor violation. If the decay
process as well as the $p\to K^+\bar{\nu}$ decay is detected in future
experiments, it may suggest the existence of sizable flavor violation in
the sfermion sector.

After all, in the presence of sfermion flavor violation, which can
naturally be sizable in the high-scale SUSY scenario, a variety of
proton decay modes may lie in a region which can be probed in future
proton decay experiments. In consequence, proton decay experiments might
shed light on SUSY even though it is broken at a relatively high-scale,
and provide a way of investigating the structure of sfermion sector.

\subsection{Flavor Constraints from Proton Decay}
\begin{figure}[t]
\begin{center}
\subfigure[Light gauginos]{\includegraphics[width=75mm]{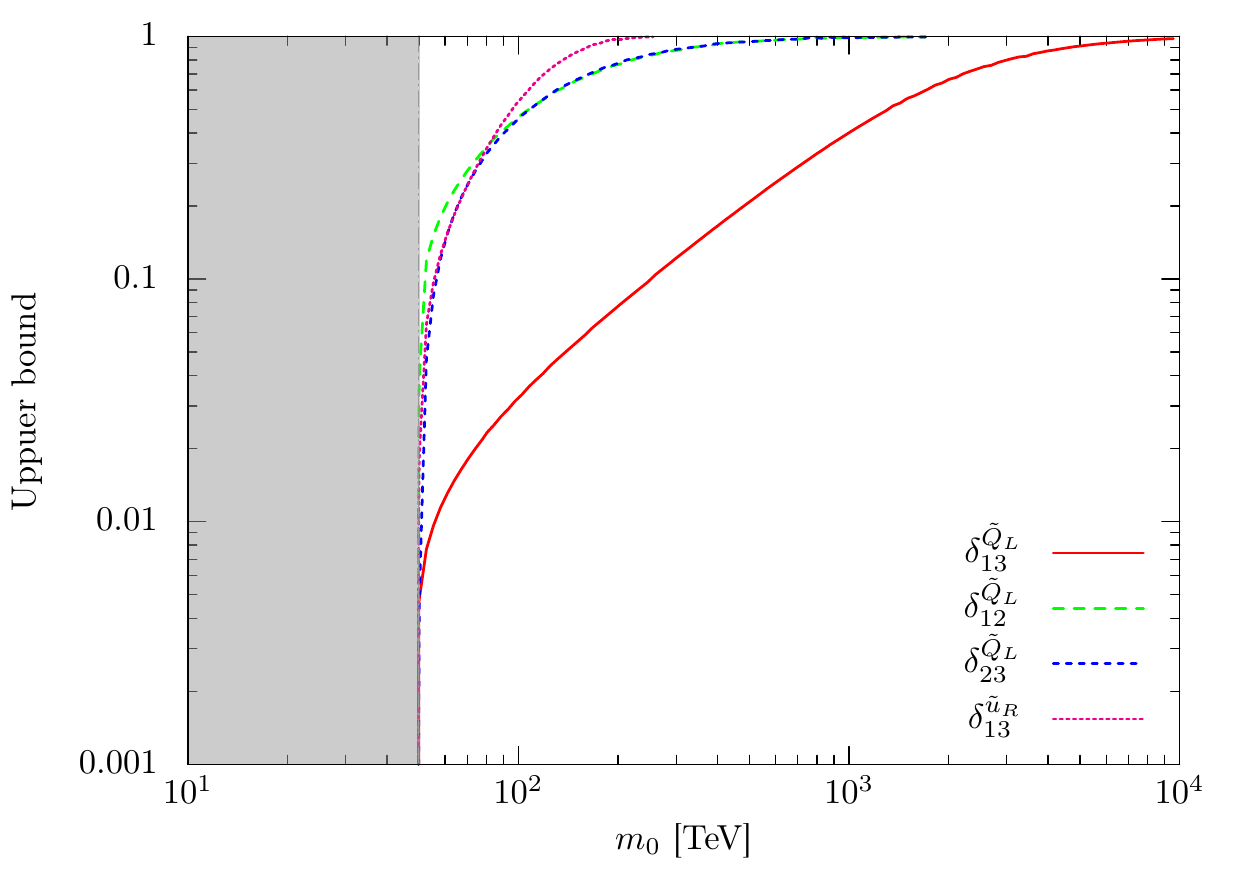}}
\subfigure[Heavy gauginos]{\includegraphics[width=75mm]{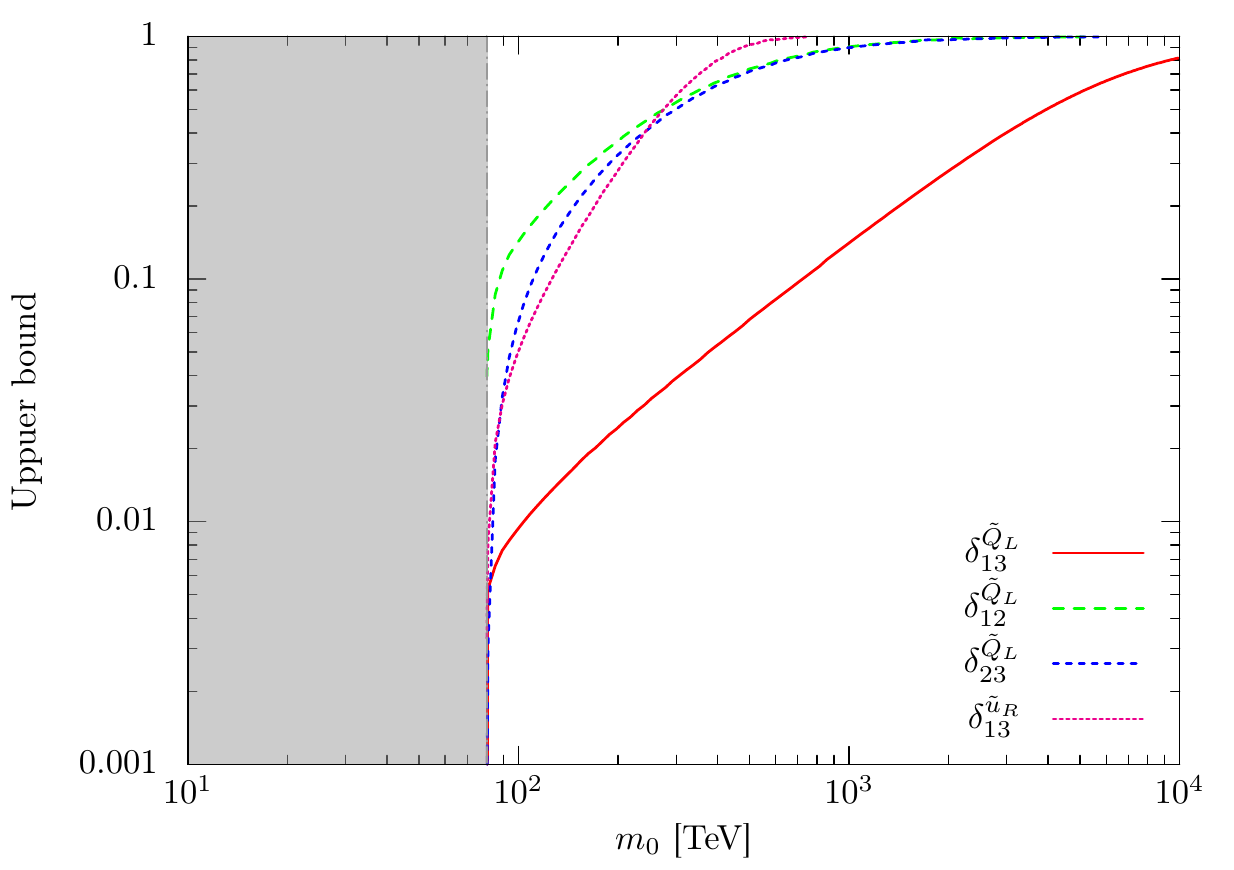}}
\caption{Upper-bound on the flavor violating mass terms $\delta$ from
 proton decay. Red, green, blue, and purple lines correspond to
 $\delta^{\widetilde{Q}_L}_{13}$, $\delta^{\widetilde{Q}_L}_{12}$,
 $\delta^{\widetilde{Q}_L}_{23}$, and  $\delta^{\widetilde{u}_R}_{13}$,
 respectively.  We take $M_{H_C} = 10^{16}$ GeV, $\mu_H = m_0$ and
 $\tan\beta = 5$. (a): $M_{\tilde B} = 600$~GeV, $M_{\tilde W} =
 300$~GeV, and  $M_{\tilde g} = - 2$~TeV. (b): $M_{\tilde B} = 6$~TeV,
 $M_{\tilde W} = 3$~TeV, and  $M_{\tilde g} = - 20$~TeV. The shaded gray
 regions show the case that the proton decay rate conflicts the current
 experimental limits, even when all $\delta$'s and $\Delta$'s are
 zero. GUT phases $P$ (defined in Eq.~\eqref{eq:GUT_CKM}) are taken so
 that the strongest bounds on $\delta$'s are obtained.
 } 
\label{fig:constraint_proton}
\end{center}
\end{figure}
As we have seen above, the sfermion flavor violations accelerate the
proton decay rate from the dimension-five operators. Therefore, in the
context of the minimal SU(5) GUT, the absence of observation of proton
decay gives constraints on the sfermion flavor violations. In
Fig.~\ref{fig:constraint_proton}, we show the upper-bound on the size of
flavor-violation $\delta$'s. Compared to the constrains from the meson
mixings (Fig.~\ref{fig:constraint}) and the EDM
(Fig.~\ref{fig:nedm}),\footnote{Notice that we expect
$\delta^{\widetilde{Q}_L}_{ij} \simeq \delta^{\widetilde{u}_R}_{ij}$ in
the minimal SU(5) GUT.  }
the proton decay stringently constrains $\delta^{\tilde Q_L}_{13}$. 
As a result, less (up-)quark EDM is predicted in the minimal SU(5) GUT.
In other words, future discovery of the quark EDM's can exclude large
parameter space of the minimal SU(5) GUT model.

\subsection{Uncertainty of Decay Rate}
Here we briefly discuss uncertainties of  estimation of the proton decay rate.
The most significant uncertainty comes from error of the hadron matrix
elements in Table~\ref{tab:matrixelements}. This provides a factor 10
uncertainty for the proton decay rate. The effects of the
experimental parameter inputs shown in Table~\ref{tab:inputs} are
relatively minor. Another important uncertainty comes from the
short-distance parameters. In addition to the color-triplet higgsino
mass $M_{H_C}$, the proton decay is quite sensitive to the Yukawa and
gauge couplings at the high-energy regions. In our analysis, however, we
do not include finite threshold effects from the sfermions and GUT
sector, and thus our result cannot achieve accuracy beyond the one-loop RGE.
To estimate possible contributions from higher order corrections we ignore,
we also study (incomplete) two-loop level RGEs. 

\begin{figure}[t]
\begin{center}
\includegraphics[width=75mm]{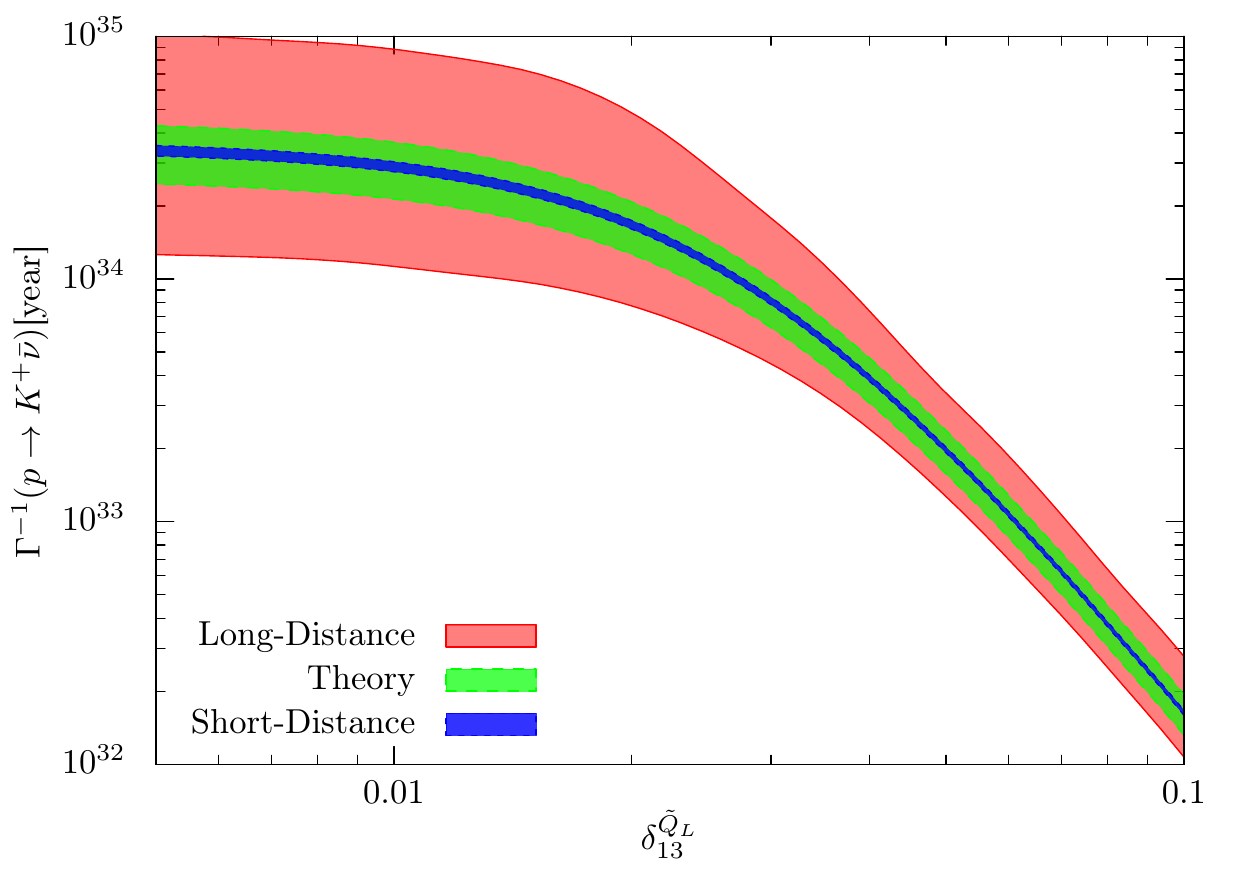}
\caption{
Error estimation of the proton decay rate. We show (one-sigma) error bands.
The SUSY mass spectrum is same as that in Fig.~\ref{fig:life_light}.
Red region displays the uncertainty from the error of the matrix
 elements shown in Table~\ref{tab:matrixelements}. Blue represents
 uncertainty from the error of the input parameters shown in Table
 \ref{tab:inputs}. Green is the theoretical uncertainties.  
}
\label{fig:error}
\end{center}
\end{figure}

In Fig.~\ref{fig:error}, we show the uncertainties in the case of $p \to
K^+\bar{\nu}$ mode. The SUSY mass spectrum is same as that in
Fig.~\ref{fig:life_light}. The red region displays the uncertainty from
the error of the matrix elements, while blue represents that from the
input parameters in Table~\ref{tab:inputs}. The green band shows the
theoretical uncertainty, which we regard as the difference between
results with the one- and two-loop RGEs.
We will discuss other contributions which may alter our present analysis
in the subsequent subsection.


\subsection{Possible Additional Corrections}

Here, we consider additional corrections which may be sizable in some
particular cases.

\subsubsection{Threshold Correction to Yukawa Couplings}

\begin{figure}[t]
\begin{center}
\includegraphics[width=70mm]{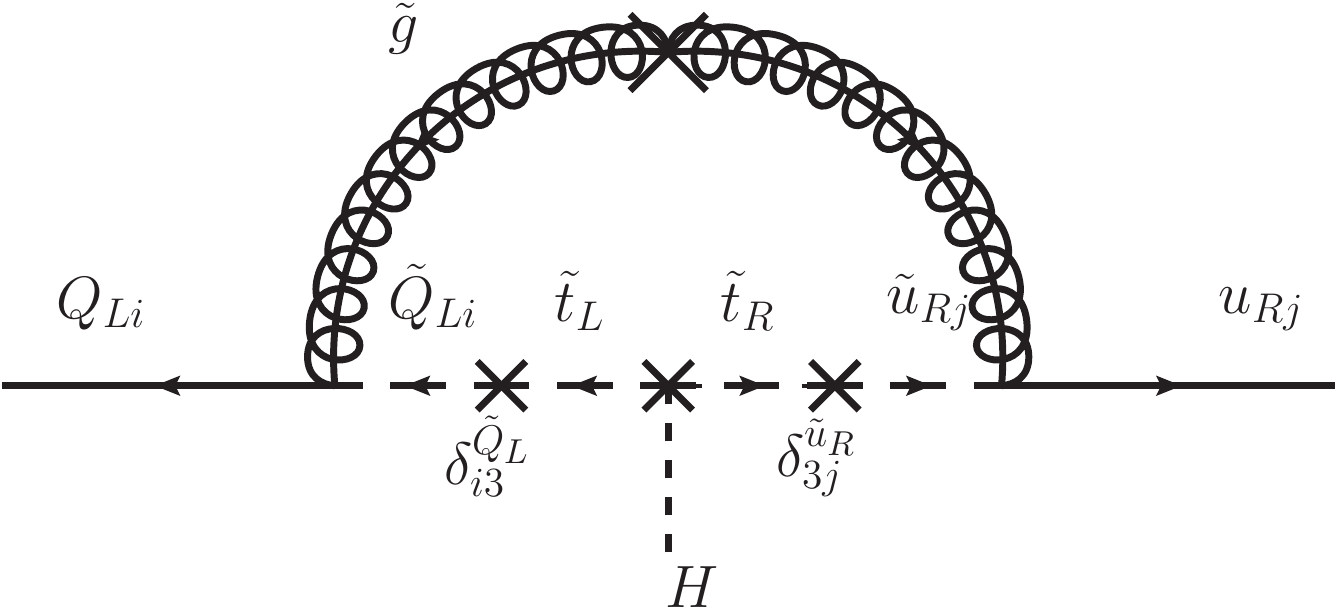}
\caption{
An example of threshold corrections to Yukawa couplings.
}
\label{fig:yukawa}
\end{center}
\end{figure}


In the present analysis, we ignore the threshold corrections to the
Yukawa couplings from sfermions as well as the GUT-scale particles or
some Planck suppressed operators. However, depending on the parameter,
these corrections may get significant. Let us first discuss the
threshold corrections at the sfermion mass scale. In
Fig.~\ref{fig:yukawa}, we show an example of such corrections. In this
case, the size of the correction is roughly given by 
\begin{equation}
\delta f^{\rm MSSM}_{ij} \sim  \frac{9}{8}\frac{f_t \alpha_3 \mu_H^*
 M^*_{\tilde g}}{4 \pi m_0^2 \tan\beta}  
\delta^{\tilde Q_L}_{i3}\delta^{\tilde u_R*}_{j3}.
 \end{equation}
Therefore, large flavor violation in the sfermion sector possibly leads
to significant corrections to the Yukawa couplings.
However note that similar processes may also give rise to EDMs in the
presence of CP violation, as discussed in Sec.~\ref{sec:edmconstraint}.
Therefore, we expect these threshold effects to be small as long as we
consider the parameter region which evades the current limits from the
EDM experiments. 

The minimal SUSY SU(5) GUT predicts the unification of down-type quark
and lepton Yukawa couplings as in Eq. (\ref{eq:unification}). However,
in the present parameter space, it is difficult to achieve the
successful Yukawa unification. This means that we omit some corrections to
the Yukawa couplings, such as those from the GUT-scale particles or some
higher-dimensional operators induced at the Planck scale. With our
ignorance of such corrections, we expect 
there is an ${\cal O}( (f_d - f_e)_{\rm GUT})$ uncertainty of estimation of 
the Yukawa couplings at the GUT scale.
It may significantly affect the prediction of the proton decay rate.
A detailed analysis will be done elsewhere \cite{NS}.

\subsubsection{Contribution from Soft Baryon-number Violating Operator}

\begin{figure}[t]
\begin{center}
\includegraphics[width=50mm]{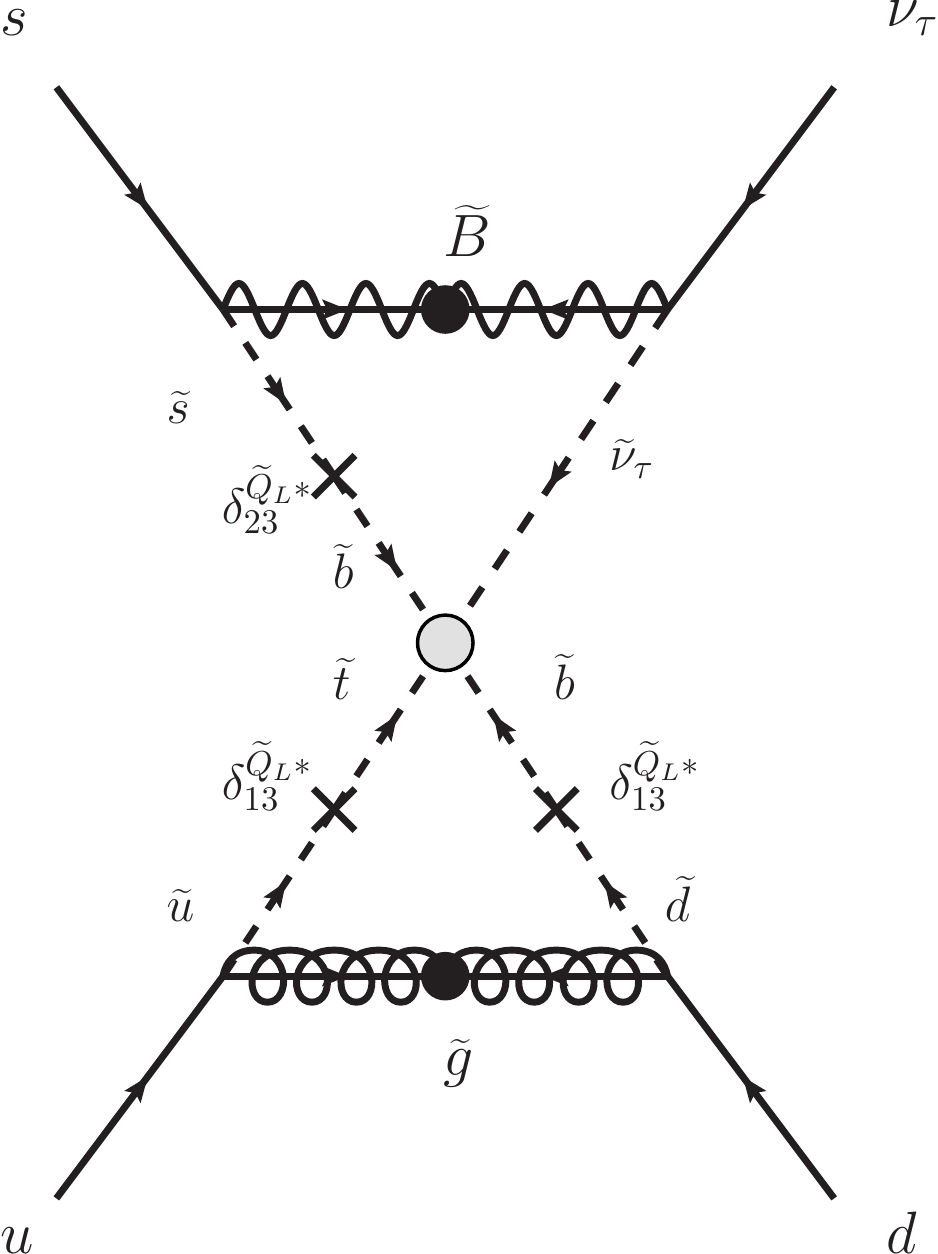}
\caption{
Contribution of the soft terms for the dimension-five operators to
 proton decay, which turns out to vanish. 
}
\label{fig:softdim5}
\end{center}
\end{figure}

Up to now, we only consider the dimension-five effective operators which
are exactly supersymmetric. However, through the supergravity effects,
the $A$-terms corresponding to these operators are also induced
\cite{Sakai:1982xn, Haba:2006dn}. This can be readily understood by
means of the superconformal compensator formalism of supergravity
\cite{Cremmer:1978hn}. In this formalism, 
the dimension-five 
operators should be accompanied by the compensator $\Sigma$ as
\begin{equation}
 \int d^2\theta ~\frac{1}{\Sigma}
\biggl[C^{ijkl}_{5L}{\cal O}_{ijkl}^{5L}+
C^{ijkl}_{5R}{\cal O}_{ijkl}^{5R}\biggr]~.
\end{equation}
Then, after the compensator gets the $F$-term VEV as $\langle \Sigma
\rangle =1+m_{3/2} \theta^2$, the dimension-four soft-terms are
induced. The leading terms are given as
\begin{equation}
{\cal L}_{\rm soft}  = -
\frac{C^L_{ijkl} m_{3/2}}{2M_{Hc}}\widetilde{Q}_{Li} \widetilde{Q}_{Lj}
\widetilde{Q}_{Lk} \widetilde{L}_{Ll} -
\frac{C^R_{ijkl} m_{3/2}}{M_{Hc}} \widetilde{u}_{Ri}^* \widetilde{d}_{Rj}^*
\widetilde{u}_{Rk}^* \widetilde{e}_{Rl}^* ~~ +{\rm h.c.}~.
\end{equation}
These soft terms also generate the proton decay four-Fermi operators via
two-loop diagrams with the exchange of gauginos and higgsinos.
This contribution is suppressed by additional factor $g^2/(16 \pi^2)
(M_{\tilde g}/m_0)$, compared to the usual one loop contribution. This
effectively results in a two-loop suppression factor in the case of
anomaly-mediation. However it is not trivial whether the $A$-term
contribution is really suppressed in the presence of large flavor
violation, since additional enhancement of the third generation Yukawa
couplings can be exploited via the flavor violation. Such an example for
the process is shown in Fig.~\ref{fig:softdim5}. To make the most of
the enhancement, all the fields included in the effective interaction
vertex, which is illustrated as a gray dot in Fig.~\ref{fig:softdim5},
should be of the third generation. Nevertheless, such a vertex is forbidden
by the antisymmetry of the color indices, and therefore the diagram
presented in Fig.~\ref{fig:softdim5} actually vanishes. After all, the
contribution of the soft terms could not use additional enhancement by
the third generation Yukawa couplings, and thus can be safely neglected
in the present calculation.


\subsubsection{$X$-Boson Contribution}
\label{sec:dim6xboson}


Next, we discuss the contribution of the SU(5) gauge boson,
$X$-boson, exchange processes to proton decay. In this case, the
effective Lagrangian is expressed in terms of the dimension-six
effective operators: 
\begin{equation}
 {\cal L}_{6}^{\rm eff}
=C_{6(1)}^{ijkl}{\cal O}^{6(1)}_{ijkl}
+C_{6(2)}^{ijkl}{\cal O}^{6(2)}_{ijkl}
~,
\end{equation}
where
\begin{align}
 {\cal O}^{6(1)}_{ijkl}&=\int d^2\theta d^2\bar{\theta}~
\epsilon_{abc}\epsilon_{\alpha\beta}
\bigl(\overline{u}^\dagger_i\bigr)^a
\bigl(\overline{d}^\dagger _j\bigr)^b
e^{-\frac{2}{3}g^\prime B}
\bigl(e^{2g_3G}Q_k^\alpha\bigr)^cL^\beta_l~,
 \\
{\cal O}^{6(2)}_{ijkl}&=\int d^2\theta d^2\bar{\theta}
\epsilon_{abc}\epsilon_{\alpha\beta}~
Q^{a\alpha}_iQ^{b\beta}_j
e^{\frac{2}{3}g^\prime B}
\bigl(e^{-2g_3G}\overline{u}^\dagger _k\bigr)^c
\overline{e}^\dagger _l~.
\end{align}
By integrating out the superheavy gauge bosons, we obtain the Wilson
coefficients as
\begin{align}
 C^{ijkl}_{6(1)}&=-\frac{g_5^2}{M_X^2}e^{i\varphi_i}\delta^{ik}
\delta^{jl}~,
\nonumber \\
 C^{ijkl}_{6(2)}&=-\frac{g_5^2}{M_X^2}e^{i\varphi_i}
\delta^{ik}(V^*)^{jl}~,
\label{eq:dim6gutmatch}
\end{align}
where $g_5$ is the unified gauge coupling constant and $M_X$ is the mass
of $X$-boson. 
Note that the results do not suffer from the model-dependence, such as the
structure of the soft SUSY breaking terms. In this sense, the SU(5)
gauge interactions provide a robust prediction for the proton decay
rate. Moreover, it is found that the resultant amplitude does not depend
on the new phases appearing in the GUT Yukawa couplings, since the factors
only affect the overall phase.

The coefficients are evolved down according to the one-loop
RGEs\footnote{The two-loop RGEs for the Wilson coefficients are also
given in Ref.~\cite{Hisano:2013ege}.},
\begin{align}
 \mu \frac{d}{d \mu}C^{ijkl}_{6(1)}&=
\biggl[
\frac{\alpha_1}{4\pi}\biggl(-\frac{11}{15}\biggr)
+\frac{\alpha_2}{4\pi}(-3)
+\frac{\alpha_3}{4\pi}\biggl(-\frac{8}{3}\biggr)
\biggr]C^{ijkl}_{6(1)}~,\nonumber \\
 \mu \frac{d}{d \mu}C^{ijkl}_{6(2)}&=
\biggl[
\frac{\alpha_1}{4\pi}\biggl(-\frac{23}{15}\biggr)
+\frac{\alpha_2}{4\pi}(-3)
+\frac{\alpha_3}{4\pi}\biggl(-\frac{8}{3}\biggr)
\biggr]C^{ijkl}_{6(2)}~,
\label{eq:rgedim61loop}
\end{align}
At the SUSY breaking scale, the
coefficients are matched with those of the four-Fermi operators as
\begin{align}
  C^{ijkl}_{(1)}(m_0)&= C^{ijkl}_{6(1)}(m_0)~, \nonumber \\
 C^{ijkl}_{(2)}(m_0)&= C^{ijkl}_{6(2)}(m_0)~.
\end{align}
The rest of the calculation is same as that carried out in
Sec.~\ref{sec:dim5formalism}. 

\begin{figure}[t]
\begin{center}
\includegraphics[height=70mm]{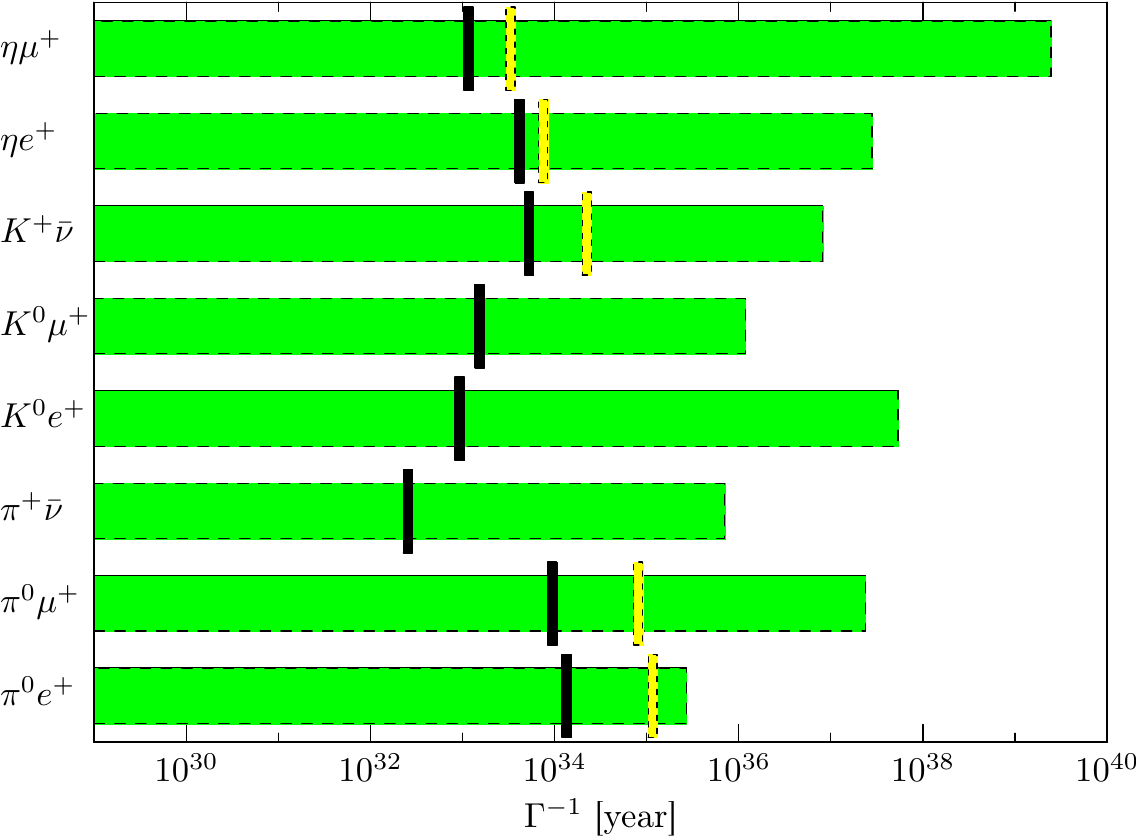}
\caption{Lifetime of each decay mode induced by the $X$-boson
 exchange. We take $m_0=100$~TeV, $M_{\widetilde{B}}=600$~GeV,
 $M_{\widetilde{W}}=300$~GeV, $M_{\widetilde{g}}=-2$~TeV, $\mu_H =m_0$,
 $M_{X}=10^{16}$~GeV, and  $\tan\beta=5$.  
Black lines represent the Super-Kamiokande constraints at 90 \%
 CL while yellow lines show the future prospects of
 Hyper-Kamiokande \cite{Shiozawa,Babu:2013jba}.}
\label{fig:chartdim6}
\end{center}
\end{figure}

Now we evaluate the decay lifetime for various modes, which are
summarized in the bar chart in Fig.~\ref{fig:chartdim6}. Here, we set
the $X$-boson mass to be $M_{X}=10^{16}$~GeV, and other parameters are
taken as follows: $m_0=100$~TeV, $M_{\widetilde{B}}=600$~GeV,
$M_{\widetilde{W}}=300$~GeV, $M_{\widetilde{g}}=-2$~TeV, $\mu_H =m_0$,
and  $\tan\beta=5$. From the figure, we see 
that the decay rates of the modes that contain different generations in
their final states are considerably suppressed, as mentioned above. This
is because in the $X$-boson exchanging process the CKM is the only
source of the flavor violation, which can be seen from
Eq.~\eqref{eq:dim6gutmatch}. Further, there is no room for the flavor
mixing effects in the sfermion mass matrices to modify the decay
rates. In this sense, the prediction given here is robust.

\section{Summary and Discussion}
\label{sec:summary}

In this paper, we have studied the impact of the sfermion flavor
structure on proton decay in the minimal SUSY SU(5) GUT model. We have
found that the flavor violation of the left-handed squark
$\delta^{\tilde Q}_{13}$ affects the proton decay rates most
significantly. The constraint on it from the proton decay bound is
stronger than that from the EDMs when the triplet Higgs mass $M_{H_C}$
is around $10^{16}$~GeV. Even if $M_{H_C} = {\cal O}(M_P)$,
$\delta^{\widetilde{Q}}_{13}$ close to unity would be confronted with the
current experimental observations.

Other mixing patterns in the left-handed squarks, as
well as those in the
right-handed up-type squarks also affect the proton decay
modes, if these $\delta$'s are close to unity. As for the other
sfermion violation, $\delta^{\tilde L_L}$, $\delta^{\tilde e_R}$ and
$\delta^{\tilde d_R}$, their impacts are small. In terms of the SU(5) GUT
matters, the flavor violation of ${\bf 10}$ matters is to be constrained
while that of $\bar{\bf 5}$ is not. This may be consistent
with observed large flavor mixing of neutrinos \cite{Hall:1999sn}. 

Further we have found that the flavor violation changes the proton decay
branch. The decay pattern of proton reflects the sfermion flavor structure.
In particular, the charged lepton modes such as $p\to \pi^0 \mu^+$ may
be smoking-gun signature of sfermion flavor violation. 
Combining indirect probes of sfermion sector via, {\it e.g.}, the
low-energy flavor and EDM measurements
\cite{McKeen:2013dma, Altmannshofer:2013lfa,Moroi:2013sfa}, gluino decay
in collider experiments \cite{Sato:2013bta,Sato:2012xf}, and
observations of gravitational waves
\cite{Saito:2012bb}, we can extract insights to the structure of
sfermion sector as well as the underlying GUT model.

We also have discussed possible corrections to the proton decay rates.
These corrections are uncertain, unless we clarify the whole picture of
the GUT model. This is beyond the scope of this paper and will be done
elsewhere \cite{NS}.

\section*{Acknowledgments}

The work of N.N. is supported by Research Fellowships of the Japan Society
for the Promotion of Science for Young Scientists. 

\section*{Appendix}
\appendix

\section{Input Parameters}
\label{sec:input}

\begin{table}[t]
\centering
 \caption{Physical parameter inputs
 \cite{UTfit,TheATLAScollaboration:2013hja,CDF:2013jga,ATLAS:2013mma,Chatrchyan:2013lba,Baak:2013ppa,Beringer:1900zz,
 Bethke:2012jm}}
\vspace{3mm}
  \label{tab:inputs}
  \begin{tabular}{|c|c|c|c|c|}
  \hline
  $m_u^{2{\rm GeV}}$ [MeV] & $m_d^{2{\rm GeV}}$ [MeV] & $m_s^{2{\rm GeV}}$ [MeV] & $m_c(m_c)$ [GeV] &$m_b(m_b)$ [GeV]\\
 \hline
    $2.15(15)$ & $4.70(20)$ & $93.5(2.5)$  & $ 1.275(25) $  &$4.18(3)$\\
  \hline \hline
   $m_t^{\rm pole}$ [GeV]&  $m_e$ [MeV] & $m_{\mu}$ [MeV] & $m_{\tau}$ [MeV] & $a_{3}(m_Z)^{5}$ \\
   \hline
     $173.24(64)$  & 0.510998918 & 105.6583692&  $1776.82(16)$
     &  $0.1184(7) $\\
       \hline \hline
      $m_h^{\rm pole}$ [GeV]& $m_W$ [GeV] &  $m_Z$ [GeV] & $(\sqrt{2} G_{\mu})^{-1/2}$ [GeV]&  \\
      \hline 
       $125.40(45)$ & $80.367(7)$ & $ 91.1875(21)$ & 246.21971 & \\
           \hline \hline
      $ \sin\theta_{12}$& $\sin\theta_{23}$ &  $\sin\theta_{13}$  & $\delta_{13}$ &  \\
      \hline 
       $0.22535(59)$ & $ 0.04173(57) $ & $ 0.00362(12)$ & $1.227(61)$ & \\
       \hline   
       
  \end{tabular}
\vspace{3mm}
\end{table}

In this section, we list the set of input parameters which we use
in our calculation. The SM parameters are summarized in
Table.~\ref{tab:inputs}. 
We take an average of the top mass measured by LHC \cite{TheATLAScollaboration:2013hja} and Tevatron \cite{CDF:2013jga} 
and the Higgs mass by ATLAS \cite{ATLAS:2013mma} and CMS \cite{Chatrchyan:2013lba}.
We adopt the fitting result of Gfitter \cite{Baak:2013ppa} as the electroweak gauge boson masses.
We use the  PDG average of the light quark masses and estimate the Yukawa couplings for the light
quarks, by using the four-loop RGEs and three-loop decoupling effects
from heavy quarks \cite{Chetyrkin:1997sg}. 
Following Ref. \cite{Buttazzo:2013uya}, we set the weak scale SM parameters.

\begin{table}[t]
\centering
\caption{Matrix elements obtained by the lattice simulation in
 Ref.~\cite{Aoki:2013yxa}.  
}
\vspace{3mm}
\label{tab:matrixelements}
\begin{tabular}{lc|lc}
\hline \hline
Matrix element&Value (GeV$^2$)& Matrix element&Value (GeV$^2$) \\[2pt]
\hline
$\langle \pi^0\vert (ud)_R^{}u_L^{}\vert p\rangle$& $-$0.103(23)(34)&
$\langle K^0\vert (us)_R ^{}u_L^{}\vert p\rangle$& 0.098(15)(12) \\
$\langle \pi^0\vert (ud)_L^{}u_L^{}\vert p\rangle$& 0.133(29)(28)&
$\langle K^0\vert (us)_L ^{}u_L^{}\vert p\rangle$& 0.042(13)(8) \\
$\langle \pi^+\vert (ud)_R^{}d_L^{}\vert p\rangle$& $-$0.146(33)(48)&
$\langle K^+\vert (us)_R ^{}d_L^{}\vert p\rangle$& $-$0.054(11)(9) \\
$\langle \pi^+\vert (ud)_L^{}d_L^{}\vert p\rangle$& 0.188(41)(40)&
$\langle K^+\vert (us)_L ^{}d_L^{}\vert p\rangle$& 0.036(12)(7) \\
$\langle \eta^0\vert (ud)_R^{}u_L^{}\vert p\rangle$& 0.015(14)(17)&
$\langle K^+\vert (ud)_R ^{}s_L^{}\vert p\rangle$& $-$0.093(24)(18) \\
$\langle \eta^0\vert (ud)_L^{}u_L^{}\vert p\rangle$& 0.088(21)(16)&
$\langle K^+\vert (ud)_L ^{}s_L^{}\vert p\rangle$& 0.111(22)(16) \\
&&
$\langle K^+\vert (ds)_R ^{}u_L^{}\vert p\rangle$& $-$0.044(12)(5) \\
&&
$\langle K^+\vert (ds)_L ^{}u_L^{}\vert p\rangle$& $-$0.076(14)(9)\\ 
\hline
\hline
\end{tabular}
\vspace{3mm}
\end{table}

We also need the hadron matrix elements for the calculation. 
In Ref.~\cite{Aoki:2013yxa}, the proton decay matrix elements are
evaluated using the direct method with $N_f=2+1$ flavor
lattice QCD, where $u$ and $d$ quarks are degenerate in mass respecting
the isospin symmetry.
The results are summarized in Table.~\ref{tab:matrixelements}. In the
table, we use an abbreviated notation like
\begin{equation}
 \langle \pi^0\vert (ud)_R^{}u_L\vert p\rangle
=
\langle \pi^0\vert\epsilon_{abc} (u^{aT}CP_Rd^b)P_Lu^c\vert p\rangle
~.
\end{equation}
The first and second parentheses represent statistical and systematic
errors, respectively. The matrix elements are evaluated at the scale of
$\mu=2$~GeV. 
In the case of the other two combinations of chirality,
the matrix elements are
derived from the above results through the parity transformation.

\section{RGEs of the Wilson Coefficients}
\label{sec:rge}

In this section, we present the RGEs for the Wilson coefficients of the
baryon-number violating operators. First, we give the RGEs of the
dimension-five proton decay operators. In this case, since the theory is
supersymmetric and the effective operators are written in terms of the
superpotential, the renormalization effects are readily
obtained from the wave-function renormalization of each chiral
superfield in the operators, thanks to the non-renormalization
theorem. We derive them at one-loop level as
\begin{align}
 \mu \frac{\partial}{\partial \mu}C^{ijkl}_{5L}(\mu)
&=\frac{1}{16\pi^2}\biggl[
\bigl(-\frac{2}{5}g_1^2-6g_2^2-8g_3^2\bigr)C^{ijkl}_{5L}
+(f_uf_u^\dagger+f_df_d^\dagger)^i_{~i^\prime} C^{i^\prime jkl}_{5L}
\nonumber \\
&+(f_uf_u^\dagger+f_df_d^\dagger)^j_{~j^\prime} C^{i j^\prime kl}_{5L}
+(f_uf_u^\dagger+f_df_d^\dagger)^k_{~k^\prime} C^{ijk^\prime l}_{5L}
+(f_ef_e^\dagger)^l_{~l^\prime} C^{ijkl^\prime}_{5L}
\biggr]~,\nonumber \\[5pt]
 \mu \frac{\partial}{\partial \mu}C^{ijkl}_{5R}(\mu)
&=\frac{1}{16\pi^2}\biggl[
\bigl(-\frac{12}{5}g_1^2-8g_3^2\bigr)C^{ijkl}_{5R}
+C^{i^\prime jkl}_{5R}(2f_u^\dagger f_u)_{i^\prime}^{~~i}\nonumber \\
&+C^{ij^\prime kl}_{5R}(2f_e^\dagger f_e)_{j^\prime}^{~~j}
+C^{ijk^\prime l}_{5R}(2f_u^\dagger f_u)_{k^\prime}^{~~k}
+C^{ijkl^\prime}_{5R}(2f_d^\dagger f_d)_{l^\prime}^{~~l}
\biggr]~.
\end{align}

Next, we evaluate the RGEs for the coefficients of the four-Fermi
operators in Eq.~\eqref{eq:fourfermidef}. We have \cite{Abbott:1980zj}
\begin{align}
 \mu \frac{\partial}{\partial \mu}C^{ijkl}_{(1)}&=
\biggl[
\frac{\alpha_1}{4\pi}\biggl(-\frac{11}{10}\biggr)
+\frac{\alpha_2}{4\pi}\biggl(-\frac{9}{2}\biggr)
+\frac{\alpha_3}{4\pi}(-4)
\biggr]C^{ijkl}_{(1)}~,\nonumber \\
 \mu \frac{\partial}{\partial \mu}C^{ijkl}_{(2)}&=
\biggl[
\frac{\alpha_1}{4\pi}\biggl(-\frac{23}{10}\biggr)
+\frac{\alpha_2}{4\pi}\biggl(-\frac{9}{2}\biggr)
+\frac{\alpha_3}{4\pi}(-4)
\biggr]C^{ijkl}_{(2)}~,\nonumber \\
 \mu \frac{\partial}{\partial \mu}C^{ijkl}_{(3)}&=
\biggl[
\frac{\alpha_1}{4\pi}\biggl(-\frac{1}{5}\biggr)
+\frac{\alpha_2}{4\pi}(-3)
+\frac{\alpha_3}{4\pi}(-4)
\biggr]C^{ijkl}_{(3)}
+\frac{\alpha_2}{4\pi}(-4)\bigl(
C^{jikl}_{(3)}+C^{kjil}_{(3)}+C^{ikjl}_{(3)}
\bigr)~,
\nonumber \\
 \mu \frac{\partial}{\partial \mu}C^{ijkl}_{(4)}&=
\biggl[
\frac{\alpha_1}{4\pi}\biggl(-\frac{6}{5}\biggr)
+\frac{\alpha_3}{4\pi}(-4)
\biggr]C^{ijkl}_{(4)}
+\frac{\alpha_1}{4\pi}(-4)C^{kjil}_{(4)}
~.
\end{align}
Here we neglect the contributions of the Yukawa couplings. In some
parameter region, inclusion of the Yukawa interaction changes the proton
decay rate by about 10~\%. Detailed analysis will be done elsewhere \cite{NS}.

Finally, we evaluate the long-distance QCD
corrections to the baryon-number violating dimension-six operators below
the electroweak scale down to the hadronic scale $\mu = 2$~GeV.
They are calculated at two-loop level in Ref.~\cite{Nihei:1994tx} as
\begin{equation}
 \mu \frac{\partial}{\partial \mu}C(\mu)=
-\biggl[
4\frac{\alpha_s}{4\pi}+\biggl(\frac{14}{3}+\frac{4}{9}N_f
+\Delta\biggr)\frac{\alpha_s^2}{(4\pi)^2}
\biggr]C(\mu)~,
\end{equation}
where $\alpha_s$ is the strong coupling constant, $N_f$ denotes the
number of quark flavors, and $\Delta=0$ ($\Delta=-10/3$) for $C_{LL}$
($C_{RL}$). The solution of the equation is
\begin{equation}
 \frac{C(\mu)}{C(\mu_0)}=\biggl[
\frac{\alpha_s(\mu)}{\alpha_s(\mu_0)}
\biggr]^{-\frac{2}{b_1}}
\biggl[
\frac{4\pi b_1+b_2\alpha_s(\mu)}
{4\pi b_1+b_2\alpha_s(\mu_0)}
\biggr]^{\bigl(\frac{2}{b_1}-\frac{42+4N_f+9\Delta}{18 b_2}\bigr)}~,
\end{equation}
with $b_1$ and $b_2$ defined by
\begin{equation}
 b_1=-\frac{11N_c-2N_f}{3}~,~~~~~
  b_2=-\frac{34}{3}N_c^2 +\frac{10}{3}N_cN_f+2C_F N_f~,
\label{b1b2}
\end{equation}
where $N_c=3$ is the number
of colors and $C_F$ is the quadratic Casimir 
invariant defined by $C_F\equiv (N_c^2-1)/{2N_c}$. By using the result,
we can readily compute the long-distance factor
\begin{equation}
 A_L\equiv \frac{C(2~{\rm GeV})}{C(m_Z)}~
\end{equation}
as follows:
\begin{align}
 A_L=\biggl[
\frac{\alpha_s(2~{\rm GeV})}{\alpha_s(m_b)}
\biggr]^{\frac{6}{25}}\biggl[
\frac{\alpha_s(m_b)}{\alpha_s(m_Z)}
\biggr]^{\frac{6}{23}}
\biggl[
\frac{\alpha_s(2~{\rm GeV})+\frac{50\pi}{77}}
{\alpha_s(m_b)+\frac{50\pi}{77}}
\biggr]^{-\frac{2047}{11550}}
\biggl[
\frac{\alpha_s(m_b)+\frac{23\pi}{29}}
{\alpha_s(m_Z)+\frac{23\pi}{29}}
\biggr]^{-\frac{1375}{8004}}~,
\end{align}
for $\Delta=0$, and
\begin{align}
 A_L=\biggl[
\frac{\alpha_s(2~{\rm GeV})}{\alpha_s(m_b)}
\biggr]^{\frac{6}{25}}\biggl[
\frac{\alpha_s(m_b)}{\alpha_s(m_Z)}
\biggr]^{\frac{6}{23}}
\biggl[
\frac{\alpha_s(2~{\rm GeV})+\frac{50\pi}{77}}
{\alpha_s(m_b)+\frac{50\pi}{77}}
\biggr]^{-\frac{173}{825}}
\biggl[
\frac{\alpha_s(m_b)+\frac{23\pi}{29}}
{\alpha_s(m_Z)+\frac{23\pi}{29}}
\biggr]^{-\frac{430}{2001}}~,
\end{align}
for $\Delta=-10/3$.
Numerically,
\begin{equation}
 A_L =
\begin{cases}
 1.257& ({\rm for}~~\Delta=0)\\
 1.253& ({\rm for}~~\Delta=-10/3)
\end{cases}
~.
\end{equation}

\section{Matching Conditions}
\label{sec:matching}

Here, we present the matching conditions for the Wilson coefficients. 

\subsection{At SUSY Breaking Scale}
\label{sec:matchingsusy}

At the sfermion mass scale, the coefficients $C_{5L}^{ijkl}$ and
$C_{5R}^{ijkl}$ for the dimension-five operators are matched to those
for the four-Fermi operators. The results are given as
\begin{align}
 C^{ijkl}_{(1)}(m_0)&=C^{ijkl}_{(1)}\vert_{\widetilde{H}}~, 
\nonumber \\[2pt]
 C^{ijkl}_{(2)}(m_0)&=C^{ijkl}_{(2)}\vert_{\widetilde{H}}~, \nonumber \\[2pt]
 C^{ijkl}_{(3)}(m_0)&=C^{ijkl}_{(3)}\vert_{\widetilde{g}}
+C^{ijkl}_{(3)}\vert_{\widetilde{W}}+C^{ijkl}_{(3)}\vert_{\widetilde{B}}
~, \nonumber \\[2pt]
 C^{ijkl}_{(4)}(m_0)&=C^{ijkl}_{(4)}\vert_{\widetilde{g}}
+C^{ijkl}_{(4)}\vert_{\widetilde{B}}
~,
\end{align}
where the subscripts $\widetilde{H}$, $\widetilde{g}$, $\widetilde{W}$,
and $\widetilde{B}$ represent the contribution of higgsino-, gluino-,
wino-, and bino-exchanging diagrams, respectively. They are computed as
follows: 
\begin{align}
 C^{ijkl}_{(1)}\vert_{\widetilde{H}}&=
\frac{1}{(4\pi)^2}(2C^{i^\prime j^\prime k l}_{5L}-C^{k i^\prime
 j^\prime l}_{5L} 
-C^{k j^\prime i^\prime  l}_{5L})
F(\mu^*_H, m_{\widetilde{Q}_I}^2, m_{\widetilde{Q}_J}^2)
\{(R^\dagger_Q)_{i^\prime I}(R_Q^{}f^*_u)_{Ii}(R^\dagger_Q) _{j^\prime J}
(R_Q^{}f_d^*)_{Jj}\}
\nonumber \\
&+\frac{1}{(4\pi)^2}(C^{*k^\prime l^\prime ij}_{5R}
-C^{*il ^\prime k^\prime j}_{5R}) F(\mu_H, m_{\widetilde{\bar u}_K}^2
, m_{\widetilde{\bar e}_L}^2)
\{
(R_{\bar{u}}^\dagger)_{k^\prime K}(R_{\bar u}^{}f_u^T)_{Kk}
(R_{\bar e}^\dagger)_{l^\prime L}(R_{\bar e}^{}f_e^T)_{Ll}
\}
~,\nonumber \\[5pt]
 C^{ijkl}_{(2)}\vert_{\widetilde{H}}&=
\frac{1}{(4\pi)^2}
(C^{ij k^\prime l^\prime}_{5L}-C^{k^\prime ji l^\prime}_{5L})
F(\mu^*_H, m_{\widetilde{Q}_K}^2, m_{\widetilde{L}_L}^2)
\{(R^\dagger_Q)_{k^\prime K}(R_Q^{}f^*_u)_{Kk}(R_L^\dagger) _{l^\prime L}
(R_L^{}f_e^*)_{Ll}\}
\nonumber \\
&+\frac{1}{(4\pi)^2}(C^{*kli^\prime j^\prime}_{5R}
-C^{*i^\prime l kj^\prime}_{5R}) F(\mu_H, m_{\widetilde{\bar u}_I}^2
, m_{\widetilde{\bar d}_J}^2)
\{
(R_{\bar u}^\dagger)_{i^\prime I}(R_{\bar u}^{}f_u^T)_{Ii}
(R_{\bar d}^\dagger)_{j^\prime J}(R_{\bar d}f_d^T)_{Jj}
\}
~.
\end{align}
\begin{align}
 C_{(3)}^{ijkl}\vert _{\widetilde{g}}=&
-\frac{4}{3}\frac{\alpha_3}{4\pi}
(C^{i^\prime j^\prime kl}_{5L}- C^{kj^\prime i^\prime l}_{5L})
F(M_{\widetilde{g}}, m^2_{\widetilde{Q}_I}, m^2_{\widetilde{Q}_J})
\bigl\{
(R^\dagger_Q)_{i^\prime I}(R_Q^{})_{Ii}(R_Q^\dagger)_{j^\prime J}(R_Q^{})_{Jj}
\bigr\}
~,\nonumber \\
 C_{(4)}^{ijkl}\vert _{\widetilde{g}}=&
-\frac{4}{3}\frac{\alpha_3}{4\pi}
\bigl[
(C_{5R}^{*i^\prime lkj^\prime}-C_{5R}^{* kl i^\prime j^\prime})
F(M_{\widetilde{g}}^*, m^2_{\widetilde{\bar u}_I}, m^2_{\widetilde{\bar d}_J})
\bigl\{
(R^\dagger_{\bar u})_{i^\prime I}(R_{\bar u}^{})_{Ii}
(R^\dagger_{\bar d})_{j^\prime J}(R_{\bar d}^{})_{Jj}
\bigr\} \nonumber \\
&-(C_{5R}^{*i^\prime lk^\prime j}-C_{5R}^{* k^\prime l i^\prime j})
F(M_{\widetilde{g}}^*, m^2_{\widetilde{\bar u}_I}, m^2_{\widetilde{\bar u}_K})
\bigl\{
(R^\dagger_{\bar u})_{i^\prime I}(R_{\bar u}^{})_{Ii}(R^\dagger_{\bar
 u})_{k^\prime K}(R_{\bar u}^{})_{Kk}
\bigr\}
\bigr]~.
\end{align}
\begin{align}
& C_{(3)}^{ijkl}\vert_{\widetilde{W}} \nonumber \\
=&\frac{\alpha_2}{4\pi}
F(M_{\widetilde{W}}, m^2_{\widetilde{Q}_I}, m^2_{\widetilde{Q}_J})
\bigl\{(R^\dagger_Q)_{i^\prime I}(R_Q^{})_{Ii}(R^\dagger_Q)_{j^\prime J}
(R_Q^{})_{Jj}\bigr\}[
(C^{i^\prime k j^\prime l}_{5L}- C^{i^\prime j^\prime kl}_{5L})
+\frac{1}{2}
(C^{ k j^\prime i^\prime l}_{5L}- C^{i^\prime j^\prime kl}_{5L})
]
\nonumber \\
+&\frac{\alpha_2}{4\pi}
F(M_{\widetilde{W}}, m^2_{\widetilde{Q}_K}, m^2_{\widetilde{L}_L})
\bigl\{(R_Q^\dagger)_{k^\prime K}(R_Q^{})_{Kk}(R^\dagger_{L})_{l^\prime L}
(R_{L}^{})_{Ll}\bigr\}[
(C^{ik^\prime  j l^\prime }_{5L}- C^{ij k^\prime l^\prime}_{5L})
+\frac{1}{2}
(C^{ k^\prime jil^\prime }_{5L}- C^{ijk^\prime l^\prime}_{5L})
]
~.
\end{align}
\begin{align}
 C_{(3)}^{ijkl}\vert_{\widetilde{B}}&=
\frac{6}{5}\frac{\alpha_1}{4\pi}[Y_{Q_L}Y_{L_L}
(C^{i j k^\prime l^\prime}_{5L}- C^{k^\prime j i l^\prime}_{5L})
F(M_{\widetilde{B}}, m^2_{\widetilde{Q}_K}, m^2_{\widetilde{L}_L})
\bigl\{
(R_Q^\dagger)_{k^\prime K}(R_Q^{})_{Kk}
(R_{L}^\dagger)_{l^\prime L}(R_{{L}}^{})_{Ll}
\bigr\}
\nonumber \\
&+Y_{Q_L}^2
(C^{i^\prime j^\prime kl}_{5L}- C^{kj^\prime i^\prime l}_{5L})
F(M_{\widetilde{B}}, m^2_{\widetilde{Q}_I}, m^2_{\widetilde{Q}_J})
\bigl\{
(R^\dagger_Q)_{i^\prime I}(R_Q^{})_{Ii}(R_Q^\dagger)_{j^\prime J}(R_Q)_{Jj}
\bigr\}
]
~,\nonumber \\
 C_{(4)}^{ijkl}\vert_{\widetilde{B}}&=
-\frac{6}{5}\frac{\alpha_1}{4\pi}
\bigl[Y_{u_R}Y_{d_R}
(C_{5R}^{*k li^\prime j^\prime}-C_{5R}^{* i^\prime l k j^\prime})
F(M_{\widetilde{B}}^*, m^2_{\widetilde{\bar u}_I}, m^2_{\widetilde{\bar
 d}_J}) \bigl\{
(R^\dagger_{\bar u})_{i^\prime I}(R_{\bar u}^{})_{Ii}(R^\dagger_{\bar
 d})_{j^\prime J}(R_{\bar d}^{})_{Jj}
\bigr\} \nonumber \\
&+Y_{u_R}^2(C_{5R}^{*i^\prime lk^\prime j}-C_{5R}^{* k^\prime l i^\prime j})
F(M_{\widetilde{B}}^*, m^2_{\widetilde{\bar u}_I}, m^2_{\widetilde{\bar u}_K})
\bigl\{
(R^\dagger_{\bar u})_{i^\prime I}(R_{\bar u}^{})_{Ii}(R^\dagger_{\bar
 u})_{k^\prime K}(R_{\bar u}^{})_{Kk}
\bigr\}\nonumber \\
&+Y_{d_R}Y_{e_R}
(C_{5R}^{*i l^\prime k j^\prime}-C_{5R}^{* k l^\prime  i j^\prime})
F(M_{\widetilde{B}}^*, m^2_{\widetilde{\bar d}_J}, m^2_{\widetilde{\bar e}_L})
\bigl\{
(R^\dagger_{\bar d})_{j^\prime J}(R_{\bar d}^{})_{Jj}(R^\dagger_{\bar
 e})_{l^\prime  L}(R_{\bar e}^{})_{Ll}
\bigr\}\nonumber \\
&+Y_{u_R}Y_{e_R}
(C_{5R}^{*k^\prime l^\prime ij}-C_{5R}^{* il^\prime  k^\prime j})
F(M_{\widetilde{B}}^*, m^2_{\widetilde{\bar u}_K}, m^2_{\widetilde{\bar
 e}_L}) \bigl\{
(R^\dagger_{\bar u})_{k^\prime K}(R_{\bar u}^{})_{Kk}(R^\dagger_{\bar
 e})_{l^\prime L}(R_{\bar e}^{})_{Ll}
\bigr\}
\bigr]~.
\label{eq:dim5gluinocontr}
\end{align}
Here, $F(M, m_1^2, m_2^2)$ is a loop-function
defined by
\begin{align}
F(M, m_1^2, m_2^2) &\equiv 
\int \frac{d^4q}{\pi^2}\frac{iM}{({q}^2-M^2)(q^2-m_1^2)(q^2-m_2^2)}
~,\nonumber \\[4pt]
&=\frac{M}{m_1^2-m_2^2}
\biggl[
\frac{m_1^2}{m_1^2-M^2}\ln \biggl(\frac{m_1^2}{M^2}\biggr)
-\frac{m_2^2}{m_2^2-M^2}\ln \biggl(\frac{m_2^2}{M^2}\biggr)
\biggr]~.
\end{align}
The matrices $R_f$ ($f=Q,L, \bar{u}, \bar{d}, \bar{e}$) are unitary
matrices which diagonalize the corresponding sfermion mass matrices; for
instance, 
\begin{align}
 R_Q^{}~ m_{\widetilde{Q}_L}^2 R_Q^\dagger&={\rm
 diag}(m^2_{\widetilde{Q}_1}, m^2_{\widetilde{Q}_2}, 
m^2_{\widetilde{Q}_3})~,\nonumber \\
 R_{\bar u}^{} (m_{\widetilde{\bar u}_R}^2)^t R_{\bar u}^\dagger&={\rm
 diag}(m^2_{\widetilde{\bar u}_1}, m^2_{\widetilde{\bar u}_2}, 
m^2_{\widetilde{\bar u}_3})~, \label{eq:mixing}
\end{align}
and so on. In the calculation, we ignore the terms suppressed by $v/m_0$
($v$ is the VEV of the Higgs field) such as the left-right mixing terms
in sfermion mass matrices.

From the above expression, it is found that in the limit of degenerate
squark masses or no flavor-mixing, the coefficients
$C_{(3)}^{ijkl}\vert_{\widetilde{g}}$ vanish; they become proportional
to $(C^{ij kl}_{5L}- C^{kj i l}_{5L})$, and thus
\begin{align}
 C_{(3)}^{ijkl}\vert_{\widetilde{g}}{\cal O}^{(3)}_{ijkl}&\propto
(C^{ij kl}_{5L}- C^{kj i l}_{5L}) {\cal O}^{(3)}_{ijkl}
=\frac{1}{2}C^{ijkl}_{5L}\bigl\{
{\cal O}^{(3)}_{ijkl}+
{\cal O}^{(3)}_{jikl}-
{\cal O}^{(3)}_{kijl}-
{\cal O}^{(3)}_{kjil}
\bigr\}
=0~.
\end{align}
The last equality immediately follows from the identity
\begin{equation}
 \epsilon^{\alpha\beta}\epsilon^{\gamma\delta}
-\epsilon^{\gamma\beta}\epsilon^{\alpha\delta}
+\epsilon ^{\alpha\gamma}\epsilon^{\delta\beta}=0~,
\end{equation}
and the Fierz identities.

In the case of $ C_{(4)}^{ijkl}\vert_{\widetilde{g}}$,
they again vanish in the degenerate mass limit. On the other hand,
they may not vanish when there is no flavor-mixing in squark mass
matrices; in this case, 
\begin{equation}
 C_{(4)}^{ijkl}\vert _{\widetilde{g}}\propto
(C^{*ilkj}_{5R}-C^{*klij}_{5R})[F(M_{\widetilde{g}},
m^2_{\widetilde{\bar u}_i},
m^2_{\widetilde{\bar d}_j})- F(M_{\widetilde{g}}, m^2_{\widetilde{\bar u}_i},
m^2_{\widetilde{\bar u}_k})]~,
\end{equation}
and thus they can remain sizable when there exists mass difference
among right-handed squarks. Their contribution to the proton decay rate
turns out to be negligible, though. Since charm quark is heavier than
proton, all we have to consider is the $i=k=1$ components, which prove
to be zero as one can see from the above expression.
Similar arguments can be applied to the case of the bino and
neutral-wino contributions. As a result, one can find that it is
the charged wino and higgsino contribution that does remain in this
limit.

\subsection{At Electroweak Scale}
\label{sec:matchingew}

Next, we give the matching conditions for the Wilson coefficients
$C_{RL}$ and $C_{LL}$ in Eq.~\eqref{efflagmz} at the electroweak scale
$\mu =m_Z$. The result is 
\begin{align}
 C_{RL}(dsu\nu_i)&=0~,\nonumber \\
 C_{RL}(usd\nu_i)&=-(V_{\rm CKM})_{j1}C^{12ji}_{(1)}(m_Z)~,\nonumber \\
 C_{RL}(uds\nu_i)&=-(V_{\rm CKM})_{j2}C^{11ji}_{(1)}(m_Z)~,\nonumber \\
 C_{LL}(dsu\nu_i)&=(V_{\rm CKM})_{j1}(V_{\rm CKM})_{k2}
[C^{jk1i}_{(3)}(m_Z)-C^{kj1i}_{(3)}(m_Z)]~,\nonumber \\
 C_{LL}(usd\nu_i)&=(V_{\rm CKM})_{j1}(V_{\rm CKM})_{k2}
C^{k1ji}_{(3)}(m_Z)~,\nonumber \\
 C_{LL}(uds\nu_i)&=(V_{\rm CKM})_{j1}(V_{\rm CKM})_{k2}
C^{j1ki}_{(3)}(m_Z)~.
\end{align}
From the equations, it is found that only the operators ${\cal
O}^{(1)}_{ijkl}$ and ${\cal O}^{(3)}_{ijkl}$
contribute to the $p\to K^+\bar{\nu}$ mode.

\section{Partial Decay Width}
\label{sec:othermode}

Here, we summarize the expressions for other decay modes than the $p\to
K^+ \bar{\nu}$ mode described in the text.

\subsection{Kaon and Charged Lepton}

The effective Lagrangian which induces the $p\to K^0 l^+_i$ ($l^+_i=e^+,
\mu^+$) mode is given as
\begin{align}
 {\cal L}(p\to K^0 l^+_i)
&=C_{RL}(usul_i)\bigl[\epsilon_{abc}(u_R^as_R^b)(u_L^cl_{Li}^{})\bigr]
+C_{LL}(usul_i)\bigl[\epsilon_{abc}(u_L^as_L^b)(u_L^cl_{Li}^{})\bigr]
\nonumber \\
&+C_{LR}(usul_i)\bigl[\epsilon_{abc}(u_L^as_L^b)(u_R^cl_{Ri}^{})\bigr]
+C_{RR}(usul_i)\bigl[\epsilon_{abc}(u_R^as_R^b)(u_R^cl_{Ri}^{})\bigr]
~.
\end{align}
The matching condition for the Wilson coefficients is
\begin{align}
 C_{RL}(usul_i)&=C^{121i}_{(1)}(m_Z)~, \nonumber \\
 C_{LR}(usul_i)&=(V_{\rm CKM})_{j2}[C^{1j1i}_{(2)}(m_Z)+
C^{j11i}_{(2)}(m_Z)]~, \nonumber \\
 C_{LL}(usul_i)&=-(V_{\rm CKM})_{j2}C^{1j1i}_{(3)}(m_Z)~, \nonumber \\
 C_{RR}(usul_i)&=C^{121i}_{(4)}(m_Z)~.
\end{align}
Then, we obtain the partial decay width as
\begin{equation}
 \Gamma (p\to  K^0 l^+_i)=
\frac{m_p}{32\pi}\biggl(1-\frac{m_K^2}{m_p^2}\biggr)^2
\bigl[
\vert {\cal A}_L(p\to K^0 l^+_i) \vert^2+
\vert {\cal A}_R(p\to K^0 l^+_i) \vert^2
\bigr]~,
\end{equation}
where
\begin{align}
 {\cal A}_L(p\to K^0 l^+_i)&=
C_{RL}(usul_i)\langle K^0\vert (us)_Ru_L\vert p\rangle
+C_{LL}(usul_i)\langle K^0\vert (us)_Lu_L\vert p\rangle~,\nonumber \\
 {\cal A}_R(p\to K^0 l^+_i)&=
C_{LR}(usul_i)\langle K^0\vert (us)_Ru_L\vert p\rangle
+C_{RR}(usul_i)\langle K^0\vert (us)_Lu_L\vert p\rangle~.
\end{align}
Notice that we have used the parity transformation to obtain the hadron
matrix elements for ${\cal A}_R$.

\subsection{Pion and Anti-neutrino}

For the $p\to \pi^+\bar{\nu}_i$ modes, the effective Lagrangian is given
as 
\begin{align}
 {\cal L}(p\to \pi^+ \bar{\nu}_i)
&=C_{RL}(udd\nu_i)\bigl[\epsilon_{abc}(u_R^ad_R^b)(d_L^c\nu_{Li}^{})\bigr]
+C_{LL}(udd\nu_i)\bigl[\epsilon_{abc}
(u_L^ad_L^b)(d_L^c\nu_{Li}^{})\bigr]
~,
\end{align}
and the matching condition for the Wilson coefficients is
\begin{align}
 C_{RL}(udd\nu_i)&=-(V_{\rm CKM})_{j1}C^{11ji}_{(1)}~,\nonumber \\
 C_{LL}(udd\nu_i)&=(V_{\rm CKM})_{j1}
(V_{\rm CKM})_{k1}C^{j1ki}_{(3)}~.
\end{align}
The partial decay width is then computed as
\begin{equation}
 \Gamma (p\to  \pi^+ \bar{\nu}_i)=
\frac{m_p}{32\pi}\biggl(1-\frac{m_\pi^2}{m_p^2}\biggr)^2
\vert {\cal A}(p\to \pi^+ \bar{\nu}_i) \vert^2~,
\end{equation}
where
\begin{align}
 {\cal A}_L(p\to \pi^+ \bar{\nu}_i)&=
C_{RL}(udd\nu_i)\langle \pi^+\vert (ud)_Rd_L\vert p\rangle
+C_{LL}(udd\nu_i)\langle \pi^+\vert (ud)_Ld_L\vert p\rangle~.
\end{align}

\subsection{Pion/eta and Charged Lepton}

The effective Lagrangian for the $p\to \pi^0 l^+_i$ is
\begin{align}
 {\cal L}(p\to \pi^0 l^+_i)
&=C_{RL}(udul_i)\bigl[\epsilon_{abc}(u_R^ad_R^b)(u_L^cl_{Li}^{})\bigr]
+C_{LL}(udul_i)\bigl[\epsilon_{abc}(u_L^ad_L^b)(u_L^cl_{Li}^{})\bigr]
\nonumber \\
&+C_{LR}(udul_i)\bigl[\epsilon_{abc}(u_L^ad_L^b)(u_R^cl_{Ri}^{})\bigr]
+C_{RR}(udul_i)\bigl[\epsilon_{abc}(u_R^ad_R^b)(u_R^cl_{Ri}^{})\bigr]
~.
\end{align}
We have the matching condition for the Wilson coefficients at the
electroweak scale as
\begin{align}
 C_{RL}(udul_i)&=C^{111i}_{(1)}(m_Z)~, \nonumber \\
 C_{LR}(udul_i)&=(V_{\rm CKM})_{j1}[C^{1j1i}_{(2)}(m_Z)+
C^{j11i}_{(2)}(m_Z)]
~, \nonumber \\
 C_{LL}(udul_i)&=-(V_{\rm CKM})_{j1}C^{1j1i}_{(3)}(m_Z)~, \nonumber \\
 C_{RR}(udul_i)&=C^{111i}_{(4)}(m_Z)~.
\end{align}
With the coefficients, the partial decay width is expressed as
\begin{equation}
 \Gamma (p\to  \pi^0 l^+_i)=
\frac{m_p}{32\pi}\biggl(1-\frac{m_\pi^2}{m_p^2}\biggr)^2
\bigl[
\vert {\cal A}_L(p\to \pi^0 l^+_i) \vert^2+
\vert {\cal A}_R(p\to \pi^0 l^+_i) \vert^2
\bigr]~,
\end{equation}
where
\begin{align}
 {\cal A}_L(p\to \pi^0 l^+_i)&=
C_{RL}(udul_i)\langle \pi^0\vert (ud)_Ru_L\vert p\rangle
+C_{LL}(udul_i)\langle \pi^0\vert (ud)_Lu_L\vert p\rangle~,\nonumber \\
 {\cal A}_R(p\to \pi^0 l^+_i)&=
C_{LR}(udul_i)\langle \pi^0\vert (ud)_Ru_L\vert p\rangle
+C_{RR}(udul_i)\langle \pi^0\vert (ud)_Lu_L\vert p\rangle~.
\end{align}

The same interaction also induces the $p\to \eta^0 l_i^+$ modes. In this
case we have
\begin{equation}
 \Gamma (p\to  \eta^0 l^+_i)=
\frac{m_p}{32\pi}\biggl(1-\frac{m_\eta^2}{m_p^2}\biggr)^2
\bigl[
\vert {\cal A}_L(p\to \eta^0 l^+_i) \vert^2+
\vert {\cal A}_R(p\to \eta^0 l^+_i) \vert^2
\bigr]~,
\end{equation}
with
\begin{align}
 {\cal A}_L(p\to \eta^0 l^+_i)&=
C_{RL}(udul_i)\langle \eta^0\vert (ud)_Ru_L\vert p\rangle
+C_{LL}(udul_i)\langle \eta^0\vert (ud)_Lu_L\vert p\rangle~,\nonumber \\
 {\cal A}_R(p\to \eta^0 l^+_i)&=
C_{LR}(udul_i)\langle \eta^0\vert (ud)_Ru_L\vert p\rangle
+C_{RR}(udul_i)\langle \eta^0\vert (ud)_Lu_L\vert p\rangle~.
\end{align}



\end{document}